\documentclass{aa}
\usepackage{graphicx}
\usepackage{natbib}
\usepackage{booktabs}
\usepackage{braket}
\usepackage{xcolor}
\usepackage{txfonts}
\newcommand{\Fermi}{\emph{Fermi}} 
\begin{document} 

\title{Ornstein-Uhlenbeck parameter extraction from light curves of \Fermi-LAT observed blazars}

\author{Paul R. Burd\inst{1},
          Luca Kohlhepp\inst{1},
          Sarah M. Wagner\inst{1},
          Karl Mannheim\inst{1}, 
          Sara Buson\inst{1}, 
          Jeffrey D. Scargle\inst{2}
          }

\institute{Institute for Theoretical Physics and Astrophysics, Julius-Maximilians-Universität Würzburg, \\
Emil-Fischer-Straße 31, Campus Hubland Nord, D-97074 Würzburg, Germany
\and 
Astrobiology and Space Science Division, NASA Ames Research Center, Moffett Field, CA 94035-1000, USA\\
Corresponding author: \email{paul.r.burd@astro.uni-wuerzburg.de}
}

\date{received 04/08/2020; accepted 22/10/2020. }
    \abstract
   {Monthly-binned $\gamma$-ray light curves of 236 bright $\gamma$-ray sources, particularly blazars, selected from a sample of 2278 high-galactic latitude objects observed with \Fermi-LAT, show  flux variability characterised by power spectral densities consisting of a single power-law component, ranging from Brownian to white noise.}
{The main goal here is to assess the Ornstein-Uhlenbeck (OU) model
by studying the range of its three parameters that reproduces these 
statistical properties.}
  {We develop procedures for extracting values of the three OU model parameters (mean flux, correlation length, and random amplitude) from time series data, and apply them to compare numerical integrations of the OU process with the \Fermi-LAT data.}
  {The OU process fully describes the statistical properties of the flux variations of the 236 blazars. The distributions of the extracted OU parameters are narrowly peaked about well-defined values 
  $(\sigma,\mu,\theta) = (0.2,-8.4,0.5)$ with variances $(0.004,0.07,0.13)$.
  The distributions of rise and decay time scales of flares in the numerical simulations, i.e. major flux variations fulfilling pre-defined criteria, 
  are in agreement with the observed ones. The power spectral densities of the synthetic light curves are statistically indistinguishable from those of the measured light curves.}
  {Long-term $\gamma$-ray flux variability of blazars on monthly time scales is well described by a stochastic model involving only three parameters.  The methods described here are powerful tools to study randomness in light curves and thereby  constrain the physical mechanisms responsible for the observed flux variations.}


    \maketitle

\section{Introduction}
The stochastic differential equation (SDE) of the Ornstein-Uhlenbeck (OU) process was originally proposed to describe the Brownian motion of particles with individual velocity $u(t)$ and average velocity $\mu$ in an ambient medium \citep{OU} .  It is given by
\begin{equation}
    \label{Eq.OU_SDE}
    \frac{\mathrm{d}u(t)}{\mathrm{d}t} = \theta (\mu - u(t)) + W(t)
\end{equation}
where $\theta$ denotes the friction parameter of the medium. It can also be interpreted as a mean reversion rate responsible for the stationarity of the process by driving the process back to the mean reversion level $\mu$. The term denoted with $W (t)$ denotes a white noise term.
Drawing identically and independently distributed (IID) values from the noise for every time step, the OU-process generates a time series.
The IID values represent causally disconnected realisations of an observable describing the physical state of the time-variable system.
Since its first use in physics, the OU process has also been employed to describe a plethora of dynamical processes in diverse fields such as economics, biology, geophysics, and in particular in astronomy. 
In each case, the three OU parameters describe similar effects for the time evolution of the processes, but must be interpreted in different ways. In economics, the model of \citet{VASICEK1977} utilizes the OU process to describe the evolution of interest rates driven by a random process, the market risk. In the context of the Vasicek model, the variable $u(t)$ in Eq. (\ref{Eq.OU_SDE}) is interpreted as the interest rate at a given time, 
$\mu$ as the long-term interest rate, $\theta$ as the speed of reversion, measuring how quick a volatile value of $u(t)$ is drawn back to the long-term interest rate, and $\sigma$ as the instant volatility, measuring at each time step the volatility of the system. 
\citet{Szarek2020} used a modification of the Vasicek model to predict the long-term price evolution of metals. In a medical study of the mechanical acceleration of humans during physical exercises \citet{2018PhyA..506..290B} used the OU process to describe how runners accelerate along the way of a certain distance in an optimum time showing that humans are able to precisely regulate their acceleration. The variable $u(t)$ in this context is interpreted as the acceleration of the runner, $\mu$ as the mean acceleration, $\theta$ as the runner's ability to vary the acceleration $u(t)$ around $\mu$, and $\sigma$ as range of the human-induced variations of the acceleration. A combination of OU-process and higher-order stochastic processes can be used to predict the breaking of an oil drill by modelling the vibrations along the drill caused by the bit-rock interaction \citep{LOBO2020}.\newline
In the context of astrophysics, studies of sources with time-variable brightness are common, typically using a multitude of discrete methods for the time-binned data.  Towards constraining physical models of the variability described by integro-differential equations, the continuous OU-process offers a promising pathway.\newline
Here, we consider the OU-process to describe the flux variations of blazars, most prominently observed at the wavelengths of $\gamma$-rays.
Blazars are a special class of highly variable astronomical sources considered to represent active galactic nuclei containing supermassive black holes which expell relativistic plasma jets producing $\gamma$-ray emission by unknown mechanisms.  So far, most physical models are constrained only by the sparse information from the spectral domain. Including the information from the time-domain offers the chance for an improvement of our understanding of these enigmatic objects. To facilitate the OU description of the variability, $u(t)$ is interpreted as a (it turns out: logarithmic) flux density value at a given time step in a time series of flux measurements, the so-called lightcurve, $\mu$ as the long-term value around which the (logarithmic) flux variations evolve\footnote{The nature of the value will be discussed later in Sec. \ref{extraction}. }, $\theta$ as the reversion rate, describing how quickly a flux value at a given time is drawn back towards $\mu$, and $\sigma$ as the so-called innovation, the amplitude of the random flux variation initiated at a given time $t$. 

 The OU-process has been employed studying the optical LCs of active galactic nuclei \citep{Kelly2009,Kelly2011,takata2018}.
Based on a small sample of 13 blazars, \citet{Sobolewska2014} showed that satisfactory fits to individual \Fermi-LAT $\gamma$-ray LCs can be obtained by simulating LCs with single or linear combinations of several OU processes. \citet{Abdo2010} found indications for a correlated noise behavior in \Fermi-LCs. The effects of the sampling time on the power spectral densities (PSDs) obtained from the LCs by a Fourier transformation were studied in \citet{timmer_koenig1995}.
Motivated by the success of simple stochastic descriptions of blazar $\gamma$-ray variability and following Ockham's principle, we question whether our sample of 236 extragalactic \Fermi-LAT LC, in particular blazars, can be reproduced employing a single (Gaussian) OU process.  Rather than fitting individual LCs, we seek to encompass the distribution of statistical properties of the observed LCs by choosing proper parameters characterizing the OU process. 

In Sec.\,\ref{Sec.sample}, the sample of \Fermi-LAT LCs (thereafter called the \emph{observed} LCs), and the data from which they were obtained, are described.  The recursion formula generating a stationary OU-process needed to obtain \emph{synthetic} LCs is derived  in Sec.\,\ref{Sec.OU}. 
In Sec.\,\ref{extraction}, we introduce our new method to extract OU parameters from time series. 
The procedure is first tested with a set of \emph{generic} time series with known OU parameters. We then extract the OU parameters of the observed LCs and generate \emph{synthetic} LCs from them. 
Section \ref{Sec.:disc} addresses to which extent the synthetic OU LCs resemble the properties of the observed \Fermi-LCs comparing their distributions of PSD slope, flux, rise and decay time, and the sparsity of large amplitude surges. 

\section{The sample of \Fermi-LAT lightcurves}
\label{Sec.sample}
In order to develop and test the methods discussed in this paper we make use of a sample of $\gamma$-ray LCs from \Fermi-LAT, which is a pair-conversion telescope sensitive to $\gamma$ rays with energies from $20\,$MeV to greater than $300\,$GeV~\citep{LATpaper:2009}. It has a large field of view ($>$ 2sr) and scans the entire sky every three hours during standard operation.
 Thanks to its almost uninterrupted all-sky monitoring since 2008, continuous observations over more than a decade are available for a large number of $\gamma$-ray sources. In this work, we utilize 28-day binned $\gamma$-ray light curves, computed at energies above 1 GeV, for 2278 extragalactic \Fermi-LAT sources. 
 In the following we briefly report the basic  procedure followed for the LAT analysis, while for more details we refer the reader to another paper were this sample was already employed \citep[see e.g.,][]{TXSMM:18}.
 The data analysis has been performed according to the \Fermi-LAT collaboration recommendations  for point-source analysis\footnote{\url{http://fermi.gsfc.nasa.gov/ssc/data/analysis/documentation/Pass8_usage.html}}.
 
LAT data of the Pass~8 source class were selected spanning the time interval from August 2008 to October 2017 and analysed using the \Fermi-LAT ScienceTools package version v11r05p3 available from the \textit{Fermi} Science Support Center\footnote{\url{http://fermi.gsfc.nasa.gov/ssc/data/analysis/}} (FSSC) and the \textsf{P8R2\_SOURCE\_V6} instrument response functions, along with the fermipy software package \citep{Wood:2017yyb}. 
 To minimize the contamination from $\gamma$-rays produced in the Earth's upper atmosphere, a zenith angle cut of $\theta<90\degr$ was applied. We applied also the standard data quality cuts \textit{($DATA\_QUAL>0$)$\&\&$($LAT\_CONFIG==1$)} and removed time periods coinciding with solar flares and $\gamma$-ray bursts detected by the LAT. 
 For each source, we selected a $10\degr \times 10\degr$ region of interest centered at its catalog position.
 Then, the $\gamma$-ray flux in each time bin was derived following a binned likelihood analysis (binned in space and energy),
 by a simultaneous fit of the source of interest and the other \Fermi-LAT sources included in a $15\degr \times 15\degr$ region, along with the Galactic and isotropic diffuse backgrounds  (gll\_iem\_ext\_v06.fits and iso\_P8R2\_SOURCE\_V6\_v06.txt).
 For all following calculations and extractions from \Fermi-LAT LCs, only flux bins with $TS\geq 9$ corresponding to a detection significance of $\sim 3 \sigma$ are considered. If more than $62\%$ of the data points in a LC are excluded, the entire LC is discarded. The threshold of $38\%$ of remaining data points has proven to yield a good balance between having enough data to perform the OU parameter extraction while excluding less significantly measured data points. This leaves 253 of the original 2278 LCs.
 Cross-checking the associations reported in the latest released \Fermi-LAT catalogs for these 253 sources, we find that there are 144  BL Lac objects, 88  FSRQs, 4  blazars (class unknown), 4 Pulsars, 4 radio galaxies, 7 unclassified sources, one AGN of unknown class, and one narrow-line Seyfert 1 Galaxy \citep{4FGL,4LAC}. In this paper we are specifically interested in blazar LCs. Therefore the OU-parameter extraction and parameter analysis is performed only for blazar type sources. This leaves 236 LCs, which will be referred to as the \emph{observed} LCs within this work. It has been shown that in order to interpret the observed LCs in a physical way, specifically in order to find the nature of the mechanism driving flux variations, time-binning on scales of minutes or possible even shorter is needed, see \citet{Shukla2020}. Such time resolution is only possible for extremely bright sources due to the necessary signal-to-noise ratio. 
 In this paper, to ensure a trade-off between the study of a sufficiently large set of LCs and  computational limitations for the analysis a large amount of \Fermi-LAT data, we chose to deploy the methods discussed in the following sections on monthly-binned LCs. This LC sample has also been used to study the long-term variability and possible periodic behaviour of the \Fermi-LAT sources, see \citet{penil2020}.
\section{Ornstein-Uhlenbeck process and stationarity}
\label{Sec.OU}
Equation (\ref{Eq.OU_SDE}) describes the OU process. The parameters and variables can be interpreted in different ways depending on the implementation of the process as a model description of a random dynamical evolution, as discussed in the introduction. In the following we discuss the conditions for the IID variables required to generate an OU process: 

\begin{subequations}
    \begin{equation}
        \label{Eq.Gamma1}
        \braket{W(t)} = 0
    \end{equation}
    \begin{equation}
        \label{Eq.Gamma2}
        \braket{W (t) \, W(t+\tau)} = \delta(\tau)
    \end{equation}
    \begin{equation}
         \label{Eq.Gamma3}
         \braket{W(t_2)|W(t_1)} = \braket{W(t_2)}
    \end{equation}
    \begin{equation}
    \label{Eq.Gamma4}
    P\left(W _1,W _2\right) = P\left(W _1\right) P\left(W _2\right)\,.
    \end{equation}
\end{subequations}
Equation (\ref{Eq.Gamma1}) requires the expectation value of $W$ to be zero, moreover Eq. (\ref{Eq.Gamma2}) requires the product of $W$ at two different time steps to average to zero, which means that $W$ is uncorrelated at any given two time steps. Equation (\ref{Eq.Gamma3}) defines $W$ as a martingale, meaning that the conditional expectation value of a given time step (given all prior time steps) is exactly the expectation value at said time step. Equation (\ref{Eq.Gamma4}) describes statistical independence: the joint probability distribution of $W$ at different time steps is required to be the product of the individual probability distributions of those. Independent draws from a normal distribution of the form $\mathcal{N}(m = 0, \sigma ^2)$ \citep{gillespie1996_OU}, for instance, fulfill Eq. (\ref{Eq.Gamma1})-(\ref{Eq.Gamma4}). The following properties of $\mathcal{N}$ are needed later on
\begin{subequations}
    \begin{equation}
    \label{Eq.N_rules1}
        \alpha + \beta \mathcal{N}\left(m,\sigma ^2\right) = \mathcal{N}\left(\alpha+\beta m, \beta ^2 \sigma ^2\right)
    \end{equation}
    \begin{equation}
        \label{Eq.N_rules2}
        \alpha + \beta \mathcal{N}(0,1) = \mathcal{N}\left(\alpha, \beta ^2\right)
    \end{equation}
    \begin{equation}
    \label{Eq.N_rules3}
        \mathcal{N}\left(m_1,\sigma_1 ^2\right)+\mathcal{N}\left(m_2,\sigma_2 ^2 \right) = \mathcal{N}\left(m_1 + m_2,\sigma _1 ^2 + \sigma _2 ^2\right) \,.
    \end{equation}
\end{subequations}
 The time differential of the white noise term can be written as
\begin{equation}
\label{Eq.time_integral_gamma}
    W (t) \mathrm{d}t = \sigma \mathcal{N}(t) \sqrt{\mathrm{d}t}
\end{equation}
with $\mathcal{N}(t) = \mathcal{N}(m =0,\sigma ^2 =1,t )$ \citep{gillespie1996_OU}. 
\subsection{Discrete updating formula}
\begin{subequations}
The time derivative in Eq. (\ref{Eq.OU_SDE}) can be rewritten as
\begin{equation}
    \frac{\mathrm{d}u}{\mathrm{d}t} = \frac{1}{\mathrm{d}t} \left( u\left(t+ \mathrm{d}t\right)-u(t)\right)
\end{equation}
solving for $u(t+\mathrm{d}t)$ and inserting Eq. (\ref{Eq.OU_SDE}) yields
\begin{equation}
    u\left(t+\mathrm{d}t\right) = u(t) +\theta \left(\mu-u(t)\right)\mathrm{d}t+W (t)\mathrm{d}t\,.
\end{equation}
In the following we chose the white noise term to be described with a Gaussian  (Eq. \ref{Eq.time_integral_gamma}), hence the SDE reads
\begin{equation}
    u\left(t+\mathrm{d}t\right) = u(t) +\theta \left(\mu-u(t)\right)\mathrm{d}t+\sigma \mathcal{N}(t)\sqrt{\mathrm{d}t} \, .    
\end{equation}

The discretisation needs to be executed such that the process equals or approximates a sampled continuous process
\begin{equation}
    u_T = u(t=\Delta t T) 
\end{equation}
where $T$ is a positive integer describing a time step. 
If the sampling of time steps is denser than the relevant time scales of the considered processes, the continuous SDE can be approximated by a discrete function $\left(\mathrm{d}t \rightarrow \Delta t \ll 1\right)$ 
\begin{equation}
    \label{Eq.Discrete_SDE}
    u_{T+1} = u_T + \theta\Delta t (\mu - u_T) + \sigma \sqrt{\Delta t} \mathcal{N}_T\,.
\end{equation}
\end{subequations}

\subsection{Stationarity}
\label{Subsec.:Stationarity}
As already mentioned in the introduction, formulae resulting from the proof of stationarity of the OU process are needed to extract parameters, $\left(\mu,\sigma,\theta\right)$ from a given LC. Proofs concerning statements that are not explicitly needed for the parameter extraction, are listed in the appendix. 
A process $X$ is stationary if its elements $x(t)$ fulfill
\begin{subequations}
    \begin{equation}
    \label{Eq.Stat1}
        \braket{x(t)} = \braket{x(t+\tau)}
    \end{equation}
    \begin{equation}
        \label{Eq.Stat2}
        \braket{x^2(t)} < \infty, \, 0 \leq t \leq \infty 
    \end{equation}
    \begin{equation}
        \label{Eq.Stat3}
        \braket{x(t)x(t+\tau)} - \braket{x(t)}\braket{x(t+\tau)} = \braket{x(\tau)x(0)}-\braket{x(\tau)}\braket{x(0)} 
    \end{equation}
    \begin{equation*}
        0 \leq t \leq \infty, \tau > 0
    \end{equation*}
\end{subequations}
Equation (\ref{Eq.Stat1}) states that the mean value is a constant function and thus independent of time, Eq. (\ref{Eq.Stat2}) ensures a finite variance for all time steps, and Eq. (\ref{Eq.Stat3}) states that the co-variance only depends on the difference between two time steps.
\subsection{Recursion Formula}
\begin{subequations}
The discrete SDE (Eq. \ref{Eq.Discrete_SDE}) is rewritten in terms of computational steps $u_{s\Delta t} = x(s)$ reads
\begin{equation}
    \label{comp_step_dOU}
    x(s) = x(s-1)+\theta\Delta t \left(\mu - x(s-1)\right) + \sigma \sqrt{\Delta t} \mathcal{N}\left(0,1,(s\Delta t)-1\right)\,.
\end{equation}
An $x(0)$ dependent explicit formula can be written as
\begin{equation}
    \begin{gathered}
    x(s) = (1-\theta \Delta t)^s x(0) +\\
    \sum _{k=1} ^s \left[(1-\theta \Delta t)^{s-k}(\theta \Delta t \mu +\sigma \sqrt{\Delta t} \mathcal{N}(0,1,(k-1)\Delta t))\right]\, ,
    \end{gathered}
\end{equation}
as shown in Appendix \ref{AppA1}. Substituting according to $n=s-k\rightarrow k=s-n$, and swapping the start and end points yields 
\begin{equation}
	\begin{gathered}
		 x(s) = (1-\theta \Delta t)^{s} x(0) +\\ \sum_{n=0}^{s-1}{[(1-\theta \Delta t)^{n} 
		\cdot(\theta \Delta t \mu + \sigma \sqrt{\Delta t} \mathcal{N}(0, 1, ((s-n)\cdot\Delta t)-1))]}.
	\end{gathered}
\end{equation}
Equation (\ref{Eq.N_rules3}) can now be applied. All normally distributed independent variables from $0$ to $((s-n)\cdot\Delta t)-1$ are summed. For each new time step a new IID variable is added. The sum equals a normal distribution at a given time step.
\begin{equation}
	\begin{aligned}
		\Rightarrow x(s) = (1-\theta \Delta t)^{s} x(0) +\\ \mathcal{N}\left(\theta \Delta t \mu \sum_{n=0}^{s-1}{(1-\theta \Delta t)^{n}}, 
		\sigma^{2}\Delta t \sum_{n=0}^{s-1}{(1-\theta \Delta t)^{2n}}, s\cdot \Delta t\right)
	\end{aligned}
\end{equation}
Using Eq. (\ref{Eq.N_rules1}) the factors can be absorbed into the normal distribution yielding
\begin{equation}
	\begin{gathered}
		\Rightarrow x(s) = \mathcal{N}((1-\theta \Delta t)^{s} x(0) + \theta \Delta t \mu \sum_{n=0}^{s-1}{(1-\theta \Delta t)^{n}}, \\
		\sigma^{2}\Delta t \sum_{n=0}^{s-1}{(1-\theta \Delta t)^{2n}}, s\cdot \Delta t)\\
		\hat{=} \mathcal{N}(\mathrm{Mean}(x),\mathrm{Var}(x),t).
		\end{gathered}
\label{Eq.x-normal}
\end{equation}
Note that Eq. (\ref{Eq.Stat1}) and (\ref{Eq.Stat2}) are fulfilled for this expression, while Eq. (\ref{Eq.Stat3}) is not. The dependence of a certain time step on any other time step vanishes when absorbing all factors into the normal distribution. The proof of the third criterion can be found in Appendix \ref{AppA2}.
\subsection{The OU process in the light of different commonly used models}

The physically motivated discrete OU model is mathematically equivalent to the autoregressive ($AR$), particularly the $AR(1)$ time series model, which in turn is equivalent to the moving average (MA) model, guaranteed by the Wold Decomposition theorem to describe any stationary  process \citep{scargle1981,scargle2020}. As \citet{scargle2020} depicts the $AR(1)$ model is typically written as
\begin{equation}
\label{AR1}
X(n) = A\,X(n-1)+R(n)\,,
\end{equation}
where $R(n)$ is uncorrelated white (not necessarily Gaussian) noise, referred to as the \textit{innovation}. Comparing Eq. (\ref{AR1}) with Eq. (\ref{comp_step_dOU}) the connection between the discrete OU process and the $AR(1)$ model becomes clear. The revision level in the $AR(1)$ model is typically set to $\mu =0$. Furthermore the $AR$ coefficient $A$ translates in the OU terminology to the terms with the mean reversion rate $(1-\theta \Delta t)$ and the innovation translates to the Gaussian white noise term $\sigma \sqrt{\Delta t} \mathcal{N}(0,1)$.\newline \citet{PKS2155} used the zero mean Continuous-Time Auto-Regressive Moving Average (CARMA(p,q)) process
\begin{equation}
\label{Eq.CARMA}
    \begin{aligned}
    \frac{\mathrm{d}^p }{\mathrm{d}t^p}y(t) + \alpha _{p-1} \frac{\mathrm{d}^{(p-1)} }{\mathrm{d}t^{(p-1)}}y(t) + ... + \alpha _0 y(t) =\\
    \beta _q \frac{\mathrm{d}^q }{\mathrm{d}t^q}\epsilon(t)+\beta _{q-1} \frac{\mathrm{d}^{(q-1)} }{\mathrm{d}t^{(q-1)}}\epsilon(t)+...+\epsilon(t)    \end{aligned}
\end{equation}
\citep{Kelly_2014_CARMA} to characterize  long- and short-term variability of the blazar PKS 2155-304. They found that the intra-night variability of this particular blazar is well described by a CARMA(1,0) process. The CARMA(1,0) process is equivalent to the Continuous-time Auto-Regressive (CAR(1)) process which in turn is also called damped random walk \citep{2014IAUS..304..395I}. CARMA(1,0)\footnote{and therefore CAR(1) and the damped random walk} is equivalent to the OU-process with $\mu = 0$. The reversion rate $\theta$ in the OU translates in the CARMA(1,0) model to $-2\alpha_0$ and the innovation is represented by the $\epsilon(t)$ term in Eq. (\ref{Eq.CARMA}). Single light curves might very well be modelled by higher order CARMA(p,q) terms, see \citet{PKS2155} or any amount of OU superpositions, see \citet{takata2018}. In this context we aim employing the simplest model to find out whether a large amount of LCs can be adequately described by a simple stochastic process and whether we are able to find a systematic clustering of the OU parameters that we wish to extract from the observed LCs. 
\end{subequations}

\section{Extraction of OU parameters from given LCs}
\label{extraction}

We refer to an OU process characterized by the parameters $(\mu,\theta,\sigma)$ from the SDE (Eq. \ref{Eq.Discrete_SDE}) as time series, which represents the logarithm (base 10) of the flux in a LC\footnote{This ensures non-negative flux densities and sparse large outbreaks as discussed in SubSec.\ref{SubSec.noneg-scarcity}}.
Based on the properties of the logarithm (base 10) of the flux, we present a mathematical description of the method to extract the parameters $\mu$, $\sigma$, and $\theta$ from any given LC. 
The procedure is realized in python code\footnote{See \url{https://github.com/PRBurd/astro-wue/tree/master/OU} for the code.}.
The parameter $\mu$ is extracted straight forward by calculating the expectation value (mean) of a given time series.
We assume that the time series can be decomposed into a white noise and a correlated colored noise part\footnote{The Wold theorem guarantees that any stationary process (e.g. a LC) can be decomposed into a deterministic and a non-deterministic part \citep{Wold1939}.}.
$\sigma$ (non-deterministic part) is obtained by finding the white noise part and the extraction of $\theta$ (deterministic part) follows from that.
We test our extraction method with a set of $10^5$ generic LCs with known input OU parameters $(\mu,\theta,\sigma)$ drawn from normal distributions of the form $(\mathcal{N}(0,1),\mathcal{N}(0,1),\mathcal{N}(5,5))$ and $\Delta t =0.1$. Unstable solutions are filtered to ensure stationarity of these generic LCs, see Eq. (\ref{Eq.var-tot-2}).
The same procedure is applied to the observed \Fermi-LCs. The resulting parameters $(\mu,\theta,\sigma)$ are then used to generate synthetic OU-LCs that mimic the observed ones.\\

\subsection{Extraction of $\mu$}
Extracting $\mu$ via the expectation value in the measured \Fermi~time series, implies that there is a state of flux which can be interpreted as stable on long time scales (with respect to the time sampling). This requires that within the detector sensitivity the source emits constantly with an underlying constant energy flux. So-called flares can then be seen as occasionally occurring (scarcity) outbreaks in the flux regime overlaying the ground state. This in turn implies that a given source for which the parameters are extracted resembles a stationary process in terms of Eqs.\ref{Eq.Stat1}-\ref{Eq.Stat3}. 

\subsection{Extraction of $\sigma$}
To extract $\sigma$, data points where $u_T$ is close to the mean (in an $\epsilon$ environment) are considered. Let $u_T$ be the sum of $\mu$ and a small deviation from $\mu$, $\epsilon$.
\begin{subequations}
    \begin{equation}
        u_T = \mu + \epsilon
    \end{equation}
the SDE (Eq. \ref{Eq.Discrete_SDE}) then reads
    \begin{equation}
        u_{T+1} = u_T + \theta \Delta t(\mu-(\mu+\epsilon)) + \sigma \sqrt{\Delta t} \mathcal{N}(t).
    \end{equation}
Solving for the difference between two sequential time steps, yields
    \begin{equation}
        u_{T+1} - u_{T} = -\theta \Delta t \epsilon + \sigma \sqrt{\Delta t} \mathcal{N}(t). 
    \end{equation}
If we chose $\epsilon \ll \frac{\sigma\sqrt{\Delta t}}{\theta \Delta t}$ the terms containing the mean reversion rate $\theta$ become negligible.
    \begin{equation}
        u_{T+1} - u_{T} = \mathcal{N}(0,\sigma ^2 \Delta t,t)
    \end{equation}
Under the condition that a given set of $u_T$ lie within the $\epsilon$ environment of $\mu$, the variance is calculated on both sides.
    \begin{equation}
    \mathrm{Var}(u_{T+1} - u_T) = \mathrm{Var}(\mathcal{N}(0,\sigma ^2 \Delta t,t)) = \sigma ^2 \Delta t
    \end{equation}
Therefore,
    \begin{equation}
        \sigma \sqrt{\Delta t} = \sqrt{\mathrm{Var}\left(\left\{u_{T+1}-u_T \middle| u_T < \epsilon \ll \frac{\sigma\sqrt{\Delta t}}{\theta \Delta t}\right\}\right)}.
    \label{Eq.extract_sigma}
    \end{equation}

\subsection{Extraction of $\theta$}
\label{subsec:extract_theta}
If the expression $\sigma \sqrt{\Delta t}$ is known, $\theta$ can be extracted from the variance of the complete time series.
With the proof of the second criterion of stationarity (see Appendix \ref{AppA3}), the variance of stationary processes reads (see Eq. \ref{Eq.var-tot-2})
    \begin{equation}
        \mathrm{Var}(u(t)) = \frac{\sigma^2 \Delta t}{1-(1-\theta\Delta t)^2}.
    \end{equation}
Solving for $\alpha = 1-\theta \Delta t$ yields
    \begin{equation}
        \alpha = \pm \sqrt{1-\frac{\sigma ^2 \Delta t}{\mathrm{Var}(u_T)}}.
        \label{Eq.alpha_pm}
    \end{equation}
The variance is insensitive to the sign of the deviation from the mean. Therefore, Eq. (\ref{Eq.alpha_pm}) yields only the absolute value for $\alpha$ reliably
    \begin{equation}
        \big|\alpha\big| = \sqrt{1-\frac{\sigma ^2 \Delta t}{\mathrm{Var}(u_T)}}
        \label{Eq.alpha_abs}
    \end{equation}

To extract the sign of $\alpha$, a similar method as in the extraction of $\sigma$ is used, however now the interest lies on data points sufficiently far from $\mu$. Due to the scarcity of relatively high values in the amplitude of LCs this method is by its very nature more imprecise than the method yielding $|\alpha|$. Nevertheless it suffices for extracting the sign. For the following consideration Eq. (\ref{Eq.Discrete_SDE}) is analyzed for large deviations of $u_T$ from $\mu$ and a small impact of $\sigma$. In this case the white noise term is small against terms dependent on the mean reversion rate $\theta$.
    \begin{equation}
        u_{T+1} = u_T +\theta \Delta t (\mu - u_T)
        \label{Eq.alpha_sign0}
    \end{equation}
with $\alpha = 1-\theta \Delta t$ and solved for $\alpha$ Eq. (\ref{Eq.alpha_sign0}) becomes 
    \begin{equation}
        \alpha_{\pm} = \left\langle \left\{\frac{u_{T+1}-\mu}{u_T -\mu} \Big{|}\mathcal{N}(t) \ll \frac{\alpha u_T + \theta \Delta t \mu}{\sigma \sqrt{\Delta t}} \right\}\right\rangle
        \label{Eq.alpha_sign1}
    \end{equation}
Equations \ref{Eq.alpha_abs} and \ref{Eq.alpha_sign1} can now be combined to obtain $\alpha$
    \begin{equation}
        \alpha = \mathrm{sign}(\alpha _{\pm}) \cdot \big|\alpha\big| 
    \end{equation}
Therefore, the mean reversion rate is
    \begin{equation}
        \theta \Delta t = 1-\alpha
    \end{equation}
    
\subsection{Constraining the $\epsilon$ parameter}
To constrain $\epsilon _\sigma$ and $\epsilon _\alpha$ defining the region close to and far from $\mu$, respectively, the relation of
    \begin{equation}
        m_\sigma = \epsilon _\sigma \,  \tilde{\sigma} _{\text{time series}}
        \label{wnr}
    \end{equation}
    \begin{equation}
        m_\alpha = \epsilon _\alpha \,  \tilde{\sigma} _{\text{time series}}
        \label{cnr}
    \end{equation}
to the standard deviation $\tilde{\sigma}$ of the time series is studied.
This is studied with a set of $10^5$ generic OU time series with parameters $(\mu,\sigma \sqrt{\Delta t},\theta \Delta t)$ drawn from normal distributions of the form $(\mathcal{N}(0,1),\mathcal{N}(0,1),\mathcal{N}(5,5))$.
The input and extracted output parameters are compared with different statistical rank correlation methods like Kendall-$\tau$, 
the Spearman and Pearson correlation coefficient and Pearson $R$ for varying $\epsilon _\sigma$ and $\epsilon _\alpha$ environments, as shown in Fig.\ref{Fig:OU_epsilon_environment_extracted}. All rank correlation methods yield consistent results for the $\epsilon _\sigma$ and $\epsilon _\alpha$ environments also showing that the results are robust for the method depending on the standard deviation of the given time series. Utilizing the standard deviation makes the method independent of $\mu$ which, if not given, would cause problems for $\mu \sim 0$. The dependence of $\epsilon$ on the standard deviation of a time series also scales with $\sigma\sqrt{\Delta t}$ and $\alpha$ which ensures that for each time series enough data points can be found within the set $\epsilon$ environments. To determine the optimum $\epsilon$ values, the gradient along the $\sigma$ axis is calculated, yielding a value for each combination of $m_{\sigma}$ and $m_{\alpha}$. The absolute values of all gradients is calculated for all $m_\sigma$ and summed over all $m_\alpha$. The optimum is defined for $m_\sigma$ that minimizes the sum. To de-couple $m_\sigma$ and $m_\alpha$
the gradients along the $m_{\alpha}$ direction are calculated for the optimum of $m_\sigma$, determined before. The minimum of the absolute gradient defines the optimum for $m_\alpha$.
Analyzing the extreme values in the 2D-histograms where the values of the rank correlation coefficients find their respective maximum, yields optimum values for the parameters in Eqs.\ref{wnr}-\ref{cnr} of $\epsilon _{\sigma} = 0.343\pm 0.0036$, resulting in $\epsilon _{\alpha} = 1.48\pm 0.36 $.
\begin{figure}
\centering
    \includegraphics[width=0.9\hsize]{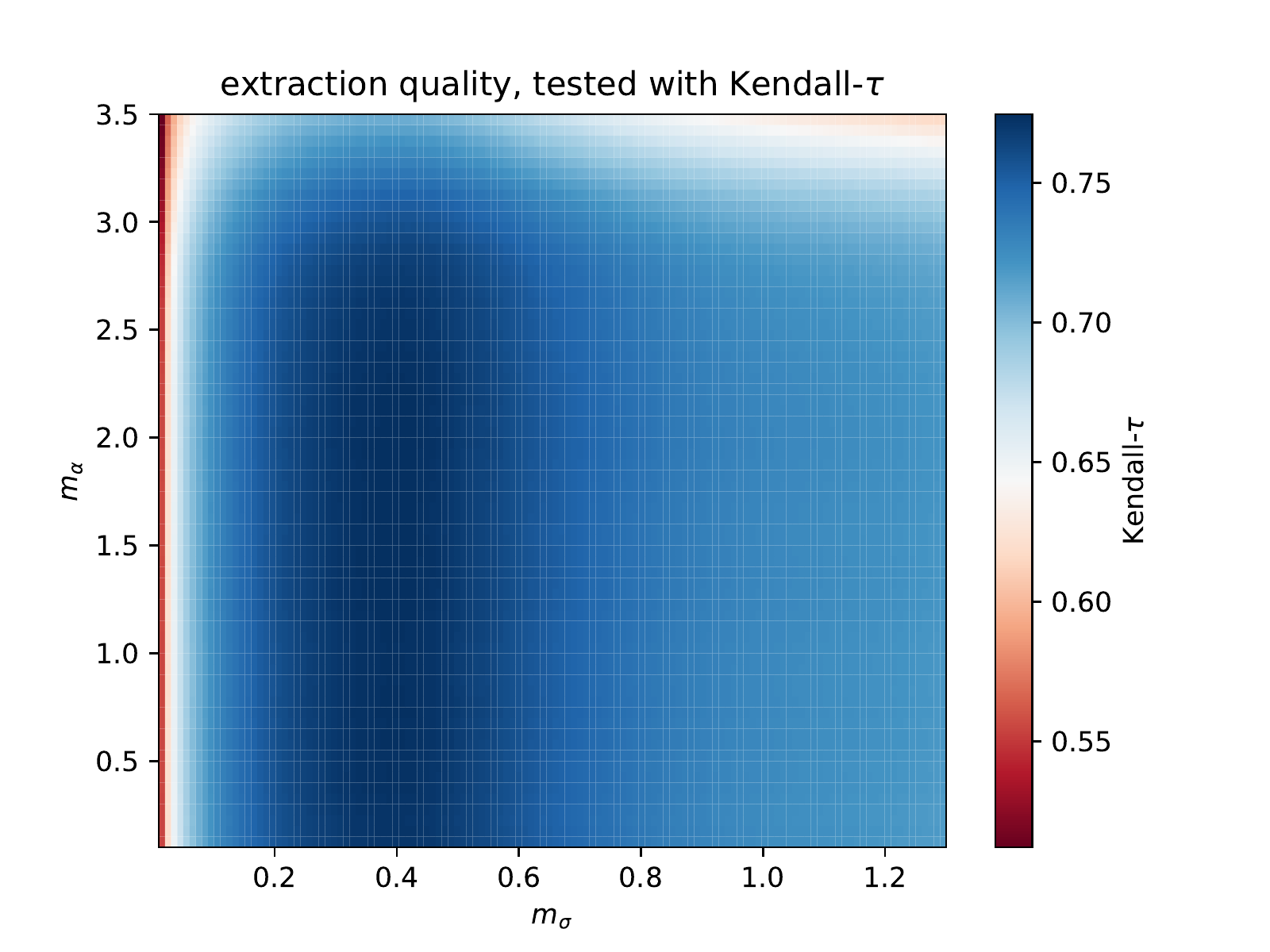}
    \includegraphics[width=0.9\hsize]{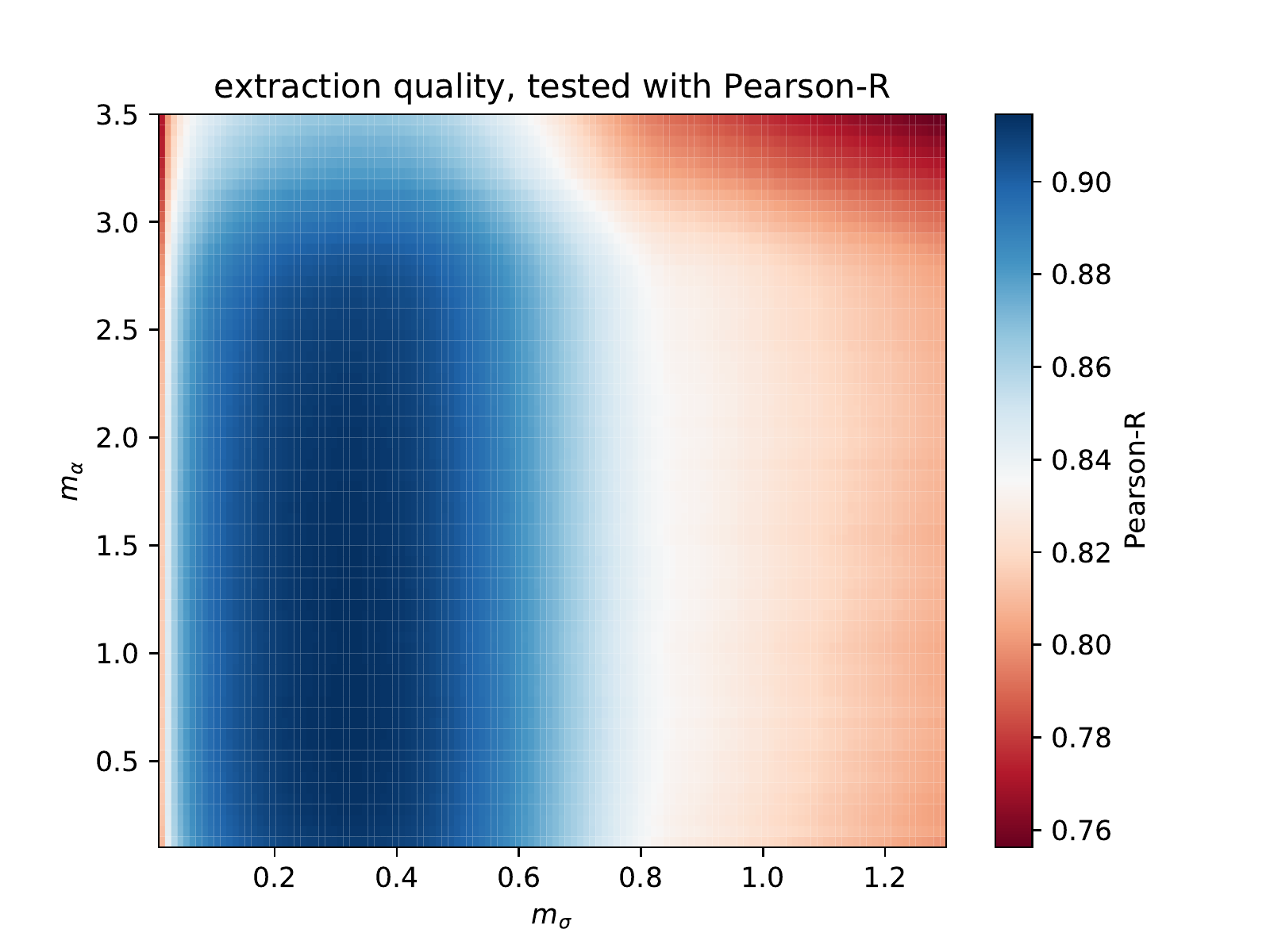}
    \includegraphics[width=0.9\hsize]{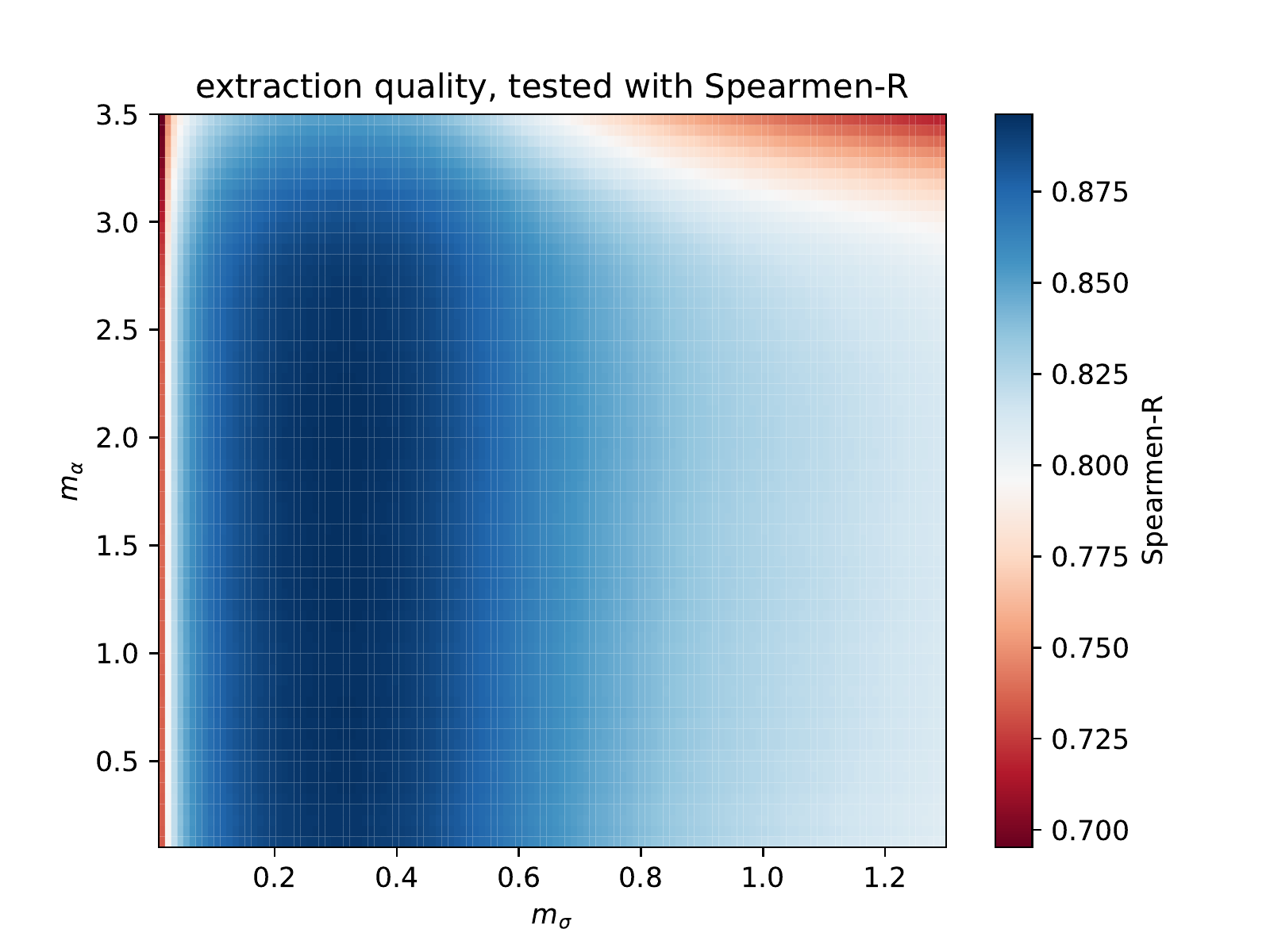}
    \caption{
    Color code shows rank correlation coefficients $\tau _{\mathrm{Kendall}}$ (top), $R_{\mathrm{Pearson}}$ (mid) and $R_{\mathrm{Spearman}}$ (bottom) between known input and extracted output OU parameters $\sigma$ and $\theta$ of the generic OU time series for different values of $m_{\sigma}$ vs $m_{\alpha}$ i.e. $\epsilon_{\sigma}$ vs $\epsilon_{\alpha}$. The blue regions define values of $\epsilon_{\sigma}$ and $\epsilon_{\alpha}$ that yield robust results for the extraction method. An analysis of the maximum valued for each rank correlation index yields environments of $\epsilon _{\sigma} = 0.343\pm 0.0036$ and $\epsilon _{\theta} = 1.48\pm 0.36 $.}
    \label{Fig:OU_epsilon_environment_extracted}
\end{figure}

\begin{figure}
\centering
  \includegraphics[width=\hsize]{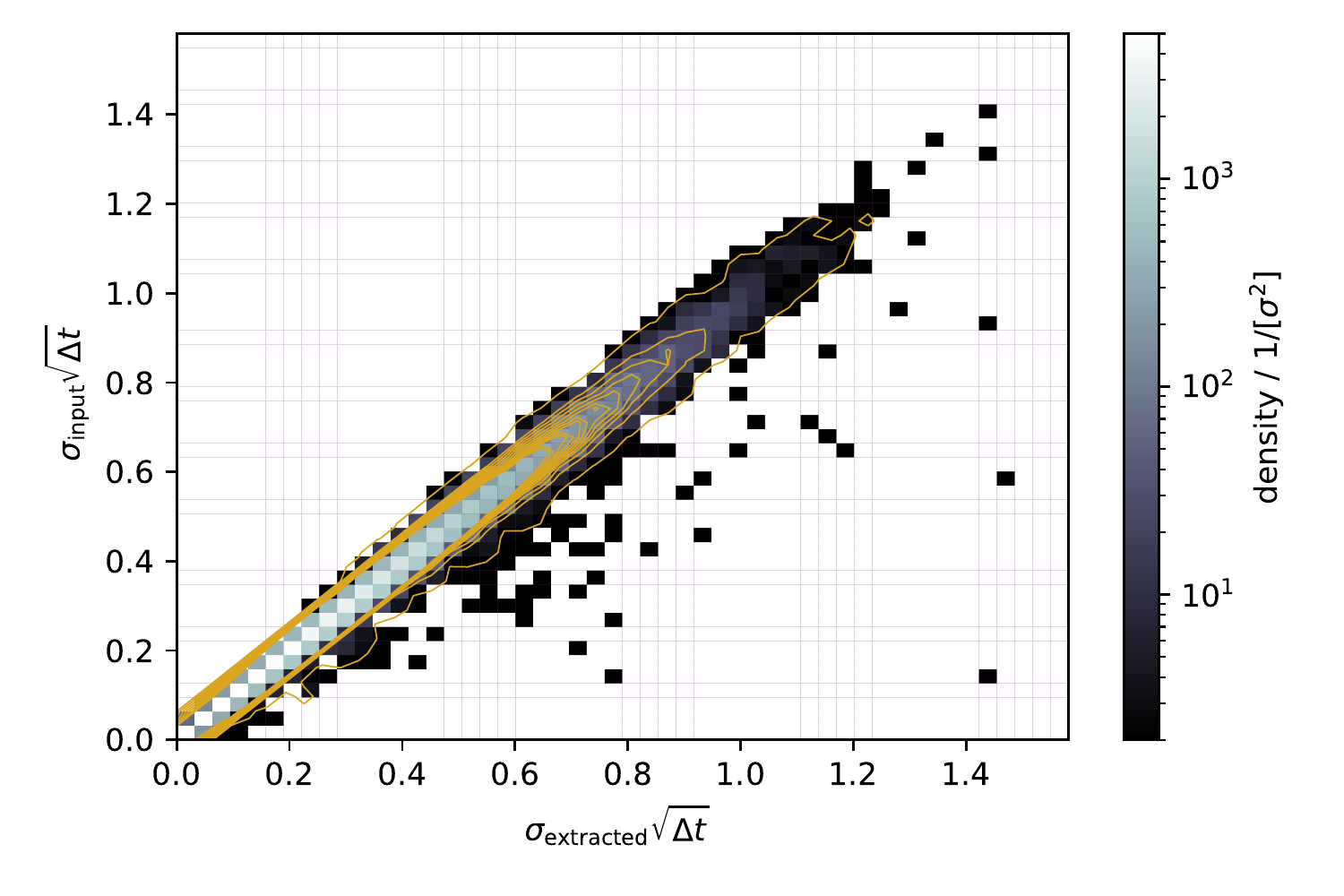}
     \caption{The 2D density histogram of the input $|\sigma|$ versus the extracted $\sigma$ shows a clear correlation. The contour lines are separated by a factor of 2 and cover a range of $[2,300]/[\sigma \sqrt{\Delta t}]$. }
       \label{Fig:OU_sigma_extracted}
\end{figure}

\begin{figure}
\centering
  \includegraphics[width=\hsize]{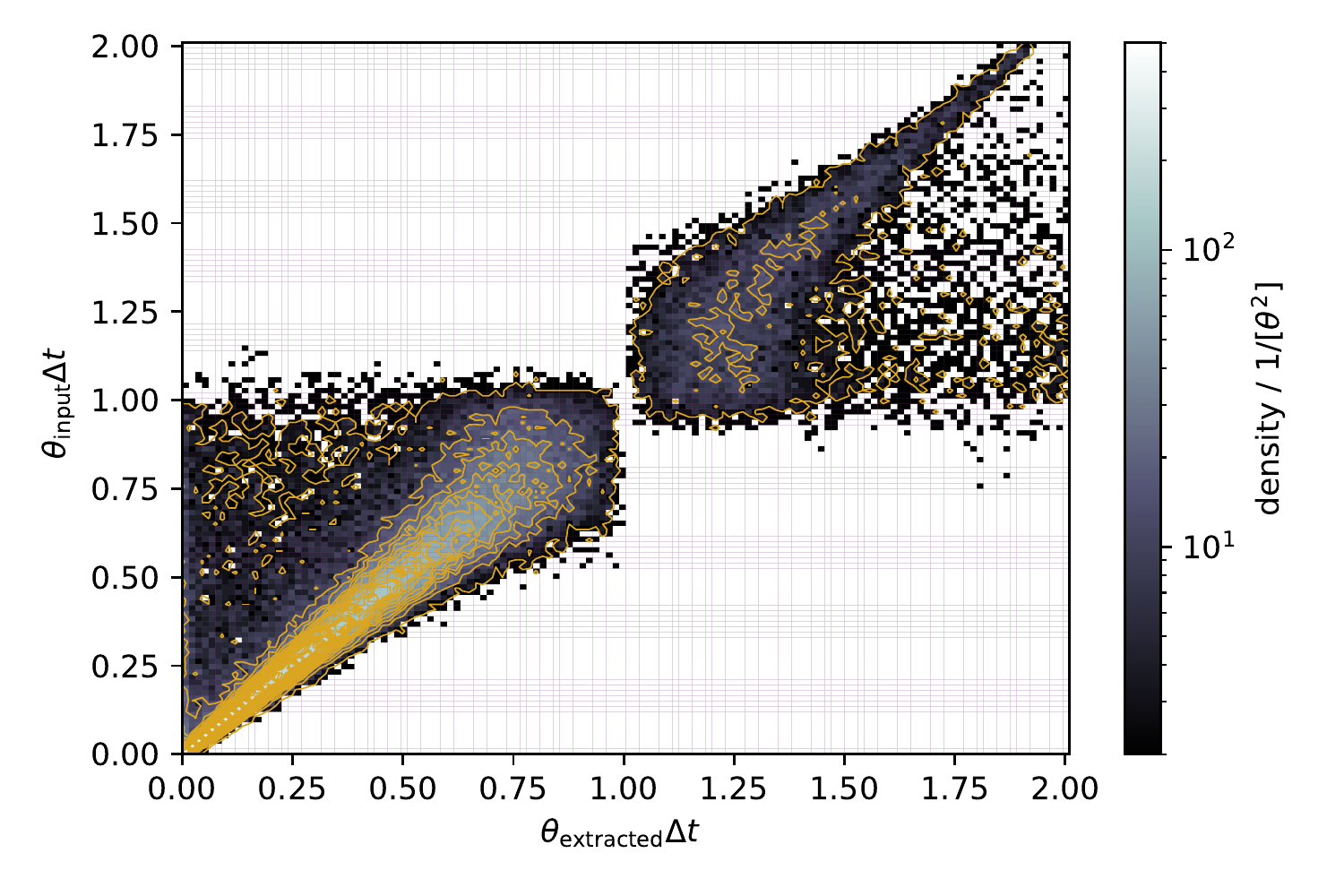}
     \caption{The inserted and extracted $\theta$ show a correlation for $\theta$ values far from one. This can be understood in the following way. If $\theta = 1$ then the reversion rate in Eq. (\ref{Eq.Discrete_SDE}) cancels leaving only the white noise term. In the case of white noise the approximation in Subsec.\ref{subsec:extract_theta} does not hold and the method breaks. The contour lines cover a range of $[2,500]/[\sigma \Delta t]$ and are separated by 48.9.}
       \label{Fig:OU_theta_extracted}
\end{figure}
\subsection{Error estimates}
To estimate the errors for the extracted $\theta \Delta t$ and $\sigma$ values, a cone is applied to the 2D histograms which includes $68\%$ of the data in Figs.\ref{Fig:OU_sigma_extracted} and \ref{Fig:OU_theta_extracted}. The opening angle in the $\sigma$ case gives an estimate for the relative error of $\Delta _\mathrm{rel} \sigma \sim 4.5\%$. The $\theta$ values yield
\begin{equation}
  \Delta \theta=\left\{
  \begin{array}{@{}ll@{}}
    17.6\% \, \theta \Delta t, & \text{if}\: 0\leq \theta \Delta t \leq 1 \\
    17.6\% \,(2-\theta \Delta t),& \text{if}\:  1<\theta \Delta t \leq 2
  \end{array}\right.
\end{equation} 
since two different cases have to be considered due to the widening of the cone up to a value of $\theta \Delta t = 1$ and it's contracting between $1< \theta \Delta t \leq 2$. The two cases arise from the fact that the reversion rate term in Eq. (\ref{Eq.Discrete_SDE}) cancels if $\theta \Delta t = 1$. In that case only the white noise term drives the time series, the approximation in Subsec.\ref{subsec:extract_theta} does not fulfill our requirements.
 With this procedure the OU parameters and their uncertainty can be extracted from any stationary time series. We apply this to the observed \Fermi-LCs and obtain 236 $\mu$, $\theta$, and $\sigma$ values that are shown in Fig.\ref{Fig:FERMI_LCs_extracted_PDFs}.

\subsection{Number generators for $\mu$, $\theta$ and $\sigma$}
\label{Subsec.:Number_generators}
In order to analyze how well the OU process can mimic the observed LCs, we generate a large sample of synthetic LCs based on the parameters extracted from the observed LCs. 
To obtain number generators returning values for $\mu$, $\sigma$, and $\theta$, the probability density functions of the extracted parameters from the LCs are fitted.
\begin{subequations}
The distributions for $\mu$ and $\theta$ are fitted with an exponentially modified Normal distribution of the form 
\begin{equation}
    f(x,\tilde{\mu},\tilde{\sigma},\lambda) = \frac{\lambda}{2} \mathrm{exp}\left(\frac{\lambda}{2}\,(2\tilde{\mu} + \lambda \tilde{\sigma} ^2 - 2x)\right) \, \mathrm{erfc}\left(\frac{\tilde{\mu}+2\tilde{\sigma} ^2 -x}{2 \sqrt{\tilde{\sigma}}}\right)\, ,
\end{equation}
see \citet{expGauss}, where $\tilde{\mu}$ is the mean of the distribution, $\tilde{\sigma}$ the variance and $\lambda = 1/\tau$ is connected to the exponential relaxation parameter $\tau$. The complementary error function is given by
\begin{equation}
    \mathrm{erfc}(x) = 1-\mathrm{erf}(x) = \frac{2}{\sqrt{\pi}} \int _x ^{\infty} e^{v^2} \mathrm{dv} \, .
\end{equation}
The fitted parameters for the $\mu$-distribution are $(\tilde{\mu},\tilde{\sigma},\lambda)_\mu \sim (-8.6,0.005,3.0)$. The the  $\theta$-distribution the fitted parameters are $(\tilde{\mu},\tilde{\sigma},\lambda)_{\theta \Delta t} \sim (0.2,0.002,3.0)$.
The $\tilde{\sigma}$ distribution is fitted with a single Gaussian of the form
\begin{equation}
    g\left(x,A,\tilde{\mu},\tilde{\sigma}\right) = \frac{A}{\tilde{\sigma} \sqrt{2\pi}} \mathrm{exp}\left(-\frac{1}{2}\left(\frac{x-\tilde{\mu}}{\tilde{\sigma}} \right)^2\right)\, .
\end{equation}
The fitted parameters for the $\sigma$ distribution are $(A,\tilde{\mu},\tilde{\sigma})_{\sigma \sqrt{\Delta t}}\sim (0.95,0.20,0.06)$. The distribution and the corresponding fits are shown in Fig.\ref{Fig:FERMI_LCs_extracted_PDFs}.
\end{subequations}
 With this, the cumulative distribution functions (CDF) can be calculated. The number generator returns a number between zero and one based on a uniform probability for each of the three parameters. To then obtain the respective parameter, the intersection of the CDF and the respective x-axis at the rolled value is determined. In this way, a set of OU parameters $\mu$, $\theta$, and $\sigma$ is obtained, given by the properties of the observed \Fermi-LCs.
For each data point calculated by the OU generator the maximum error can then be estimated to $\Delta _{\mathrm{rel}} \mathrm{OU} _i \sim 18.1\%$ based on the error cones discussed above.

\begin{figure}
\centering
    \includegraphics[width=\hsize]{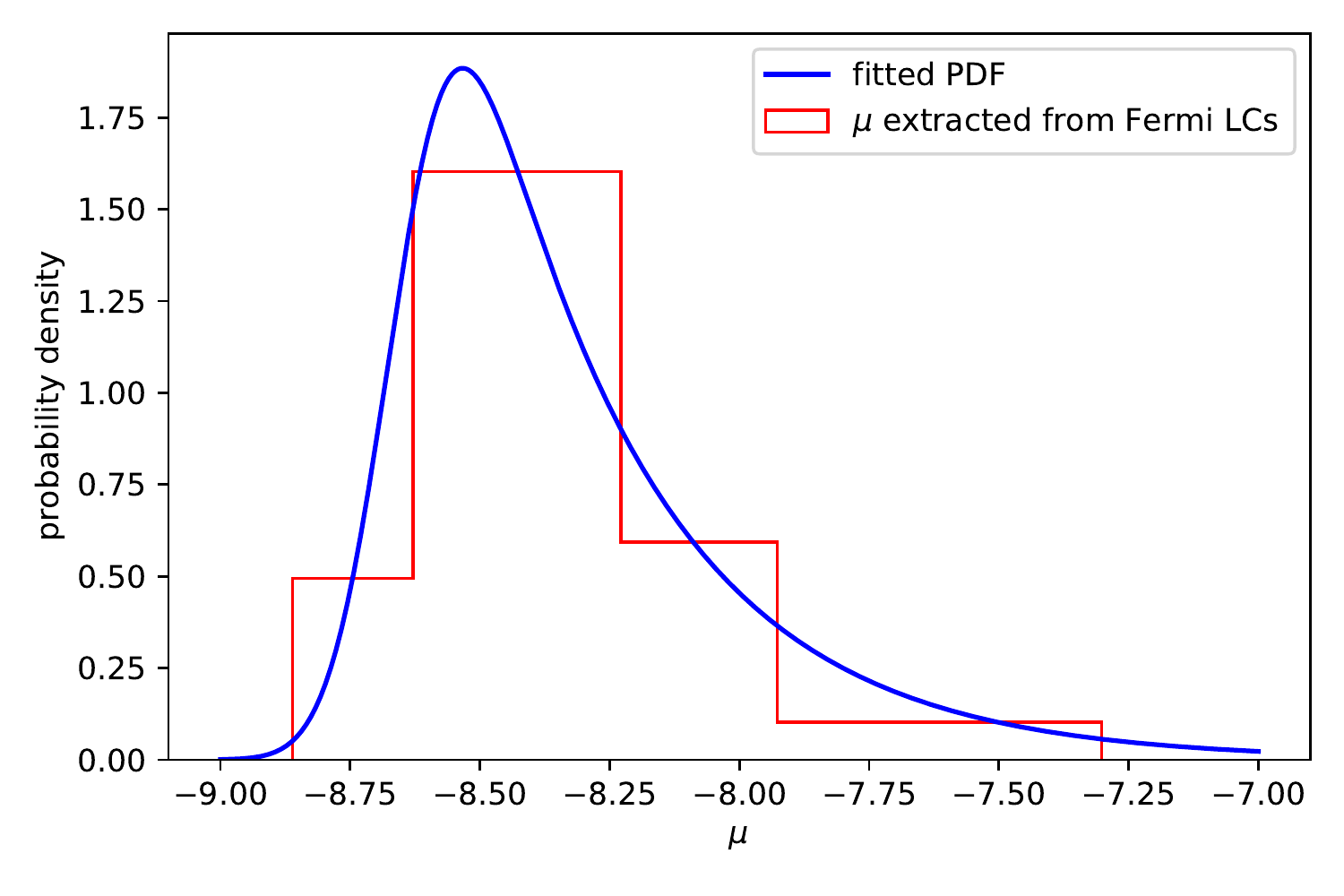}
    \includegraphics[width=\hsize]{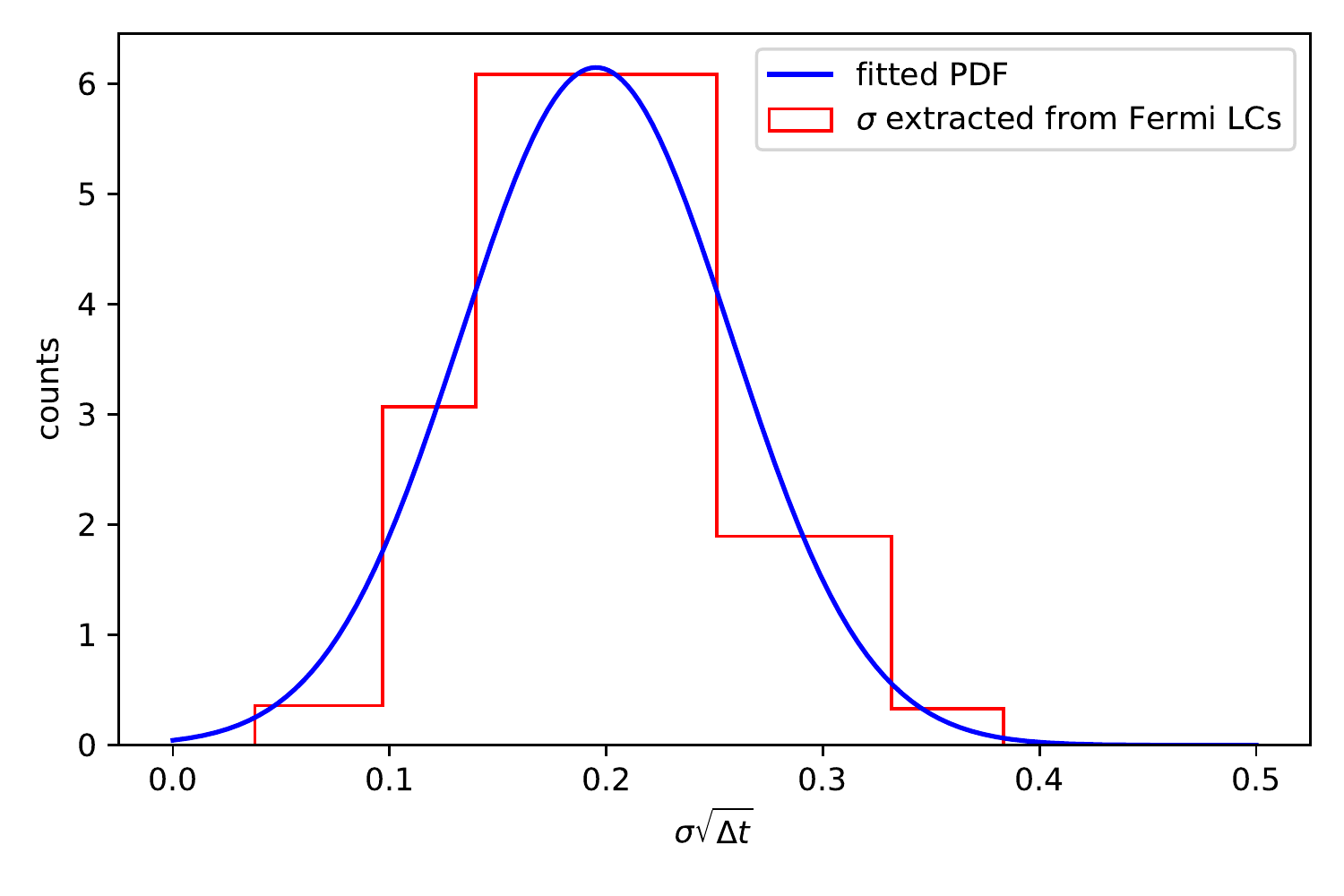}
    \includegraphics[width=\hsize]{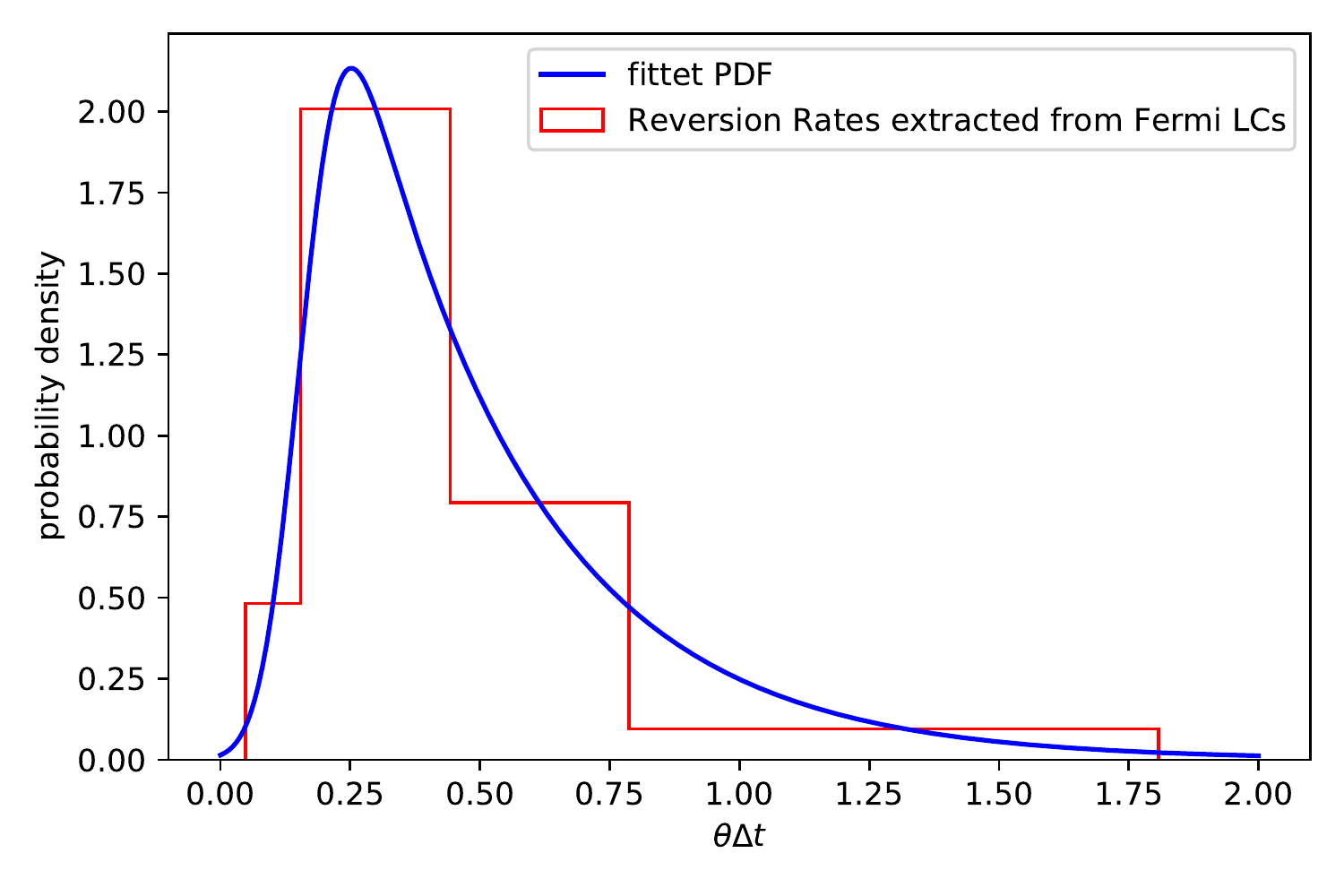}
    \caption{Extracted parameters from the \Fermi~time series with the fitted PDFs (singlea Gaussian for $\mu$ and $\sigma$, two Gaussians for $\theta$). \textit{Top}: Extracted expectation values $\mu$ of the \Fermi~time series. \textit{Mid}: Extracted white noise prominence of the \Fermi~time series $\sigma \sqrt{\Delta t}$. \textit{Bottom}: Extracted reversion rate of the \Fermi~time series $\theta\Delta t$.  }
    \label{Fig:FERMI_LCs_extracted_PDFs}
\end{figure}


\end{subequations}
\section{Results and discussion}
\label{Sec.:disc}
We simulated $10^5$ OU time series with parameters $(\mu,\sigma,\theta)$ drawn from the number generators described in Sect. \ref{Subsec.:Number_generators}.   To discuss their implications for the interpretation of the observed LCs, we obtained the
innovation as the product of the drawn $\sigma$ value and the standard normal distribution with a variance of 1 and centered around 0, $\left[\mathcal{N}(0,1)\right]$ and transformed the simulated time series to synthetic flux LCs retaining their scarcity of large flux variations and non-negative flux values.
\subsection{Scarcity, non-negativity and flux distributions}
\label{SubSec.noneg-scarcity}
\begin{figure}
\centering
  \includegraphics[width=\hsize]{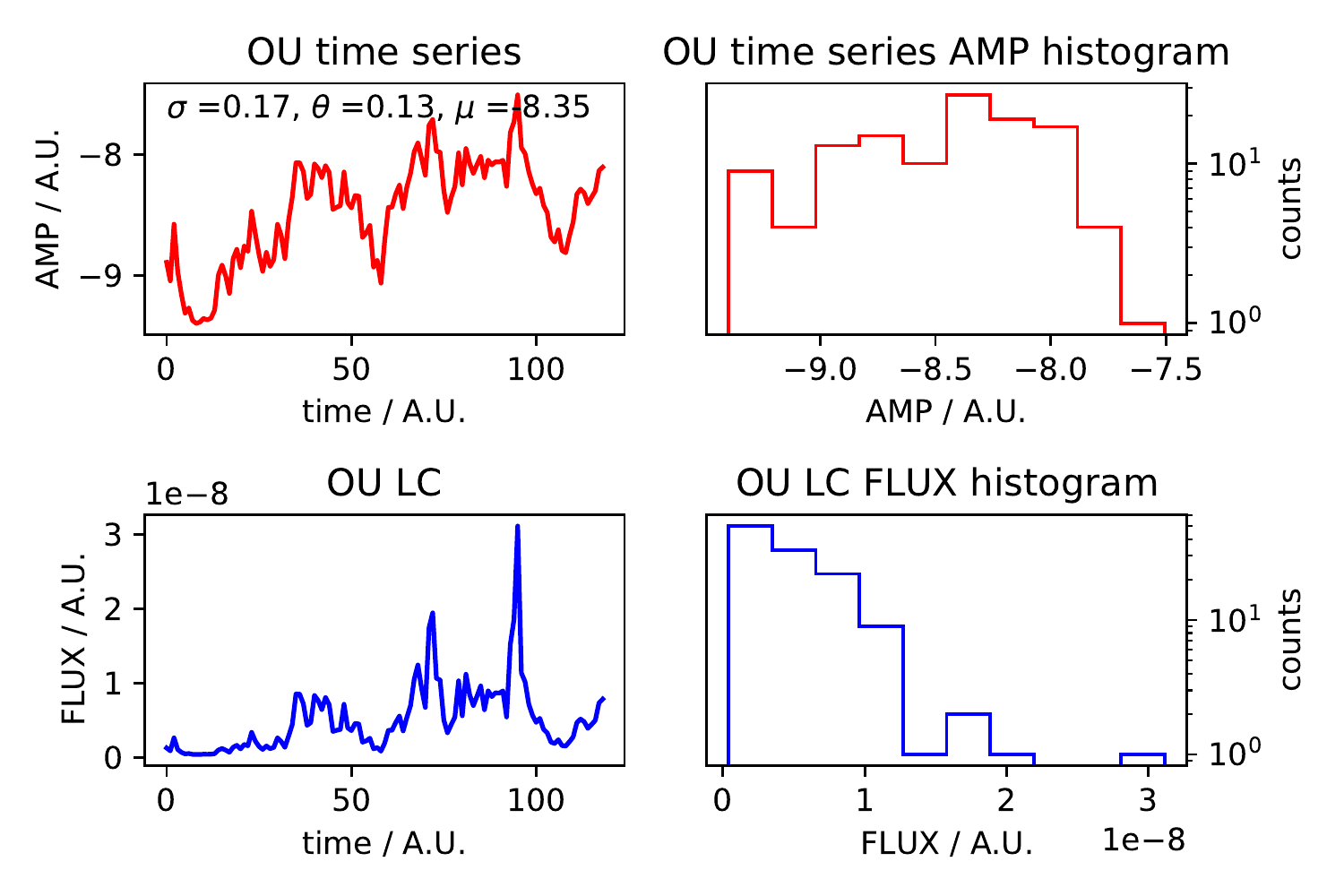}
      \caption{The OU time series and the OU light curves (left) are shown with the respective flux distributions (right). Top: Generated OU time series. Bottom: OU light curve.}
    \label{Fig:OU_scarce_TSLC}
\end{figure}

For the discussion the terms \textit{time series} and \textit{light curve} are connected in the following way. The OU process itself yields a time series $X_{OU}$, the corresponding OU LC $Y_{OU}$ is then given by  
\begin{subequations}
    \begin{equation}
    Y_{OU} = 10^{X_{OU}}
    \end{equation}
The same logic is applied to \Fermi-LAT LCs. The \Fermi-LAT LC $Y_F$ is connected to its time series $X_F$ by
    \begin{equation}
    X_F = \mathrm{log}_{10} (Y_F)    
    \end{equation}
\end{subequations}
A time series obtained with the OU generator yields a synthetic data stream scattered around an expectation value. One instance of such a synthetic time series is shown in Fig.\ref{Fig:OU_scarce_TSLC} on the top right with the corresponding generated OU parameters.
The relatively low mean reversion rate allows the process to evolve to greater deviations from the expectation value before being forced back toward the mean. This has implications on the final flux distribution of the OU LCs. \newline
To generate the OU LC, it is important to transform the time series in a way that the values are non-negative and a certain scarcity is given, reflecting the fact that most sources show strong outbursts only relatively seldom with respect to a mean flux. If the source driving the white noise part of the OU process is to be a physical source of radiation it has to yield non-negative flux densities. In addition, a Gaussian process in itself does not produce a sufficient scarcity of large outliers and can therefore not be a good model for a flare process. \citet{scargle2020} shows that it is possible to achieve non-negativity and scarcity by defining the white noise term in a way where the auto-regressive process itself results in a time series with these attributes. In this paper, however, rather than tweaking the white noise term in a special way, we obtain the OU light curve by taking the time series to a power of ten.  If the OU process is considered as a model of the flux variability, the exponentiation is motivated by the idea that random fluctuations of plasma properties have to first grow exponentially to give rise to the observed $\gamma$-ray variability, as we are dealing with non-thermal processes far from equilibrium.  Another, more specific motivation, is to consider differential Doppler boosting to be responsible for the flux variations.
In this case, the exponentiation results from the relativistic beaming of particles radiating primarily into the direction of motion, as the direction of this motion sweeps across the line of sight to the observer \citep{cohen2007}.

\subsection{Power spectral density}
\begin{figure*}
\centering
    \includegraphics[width=0.45\hsize]{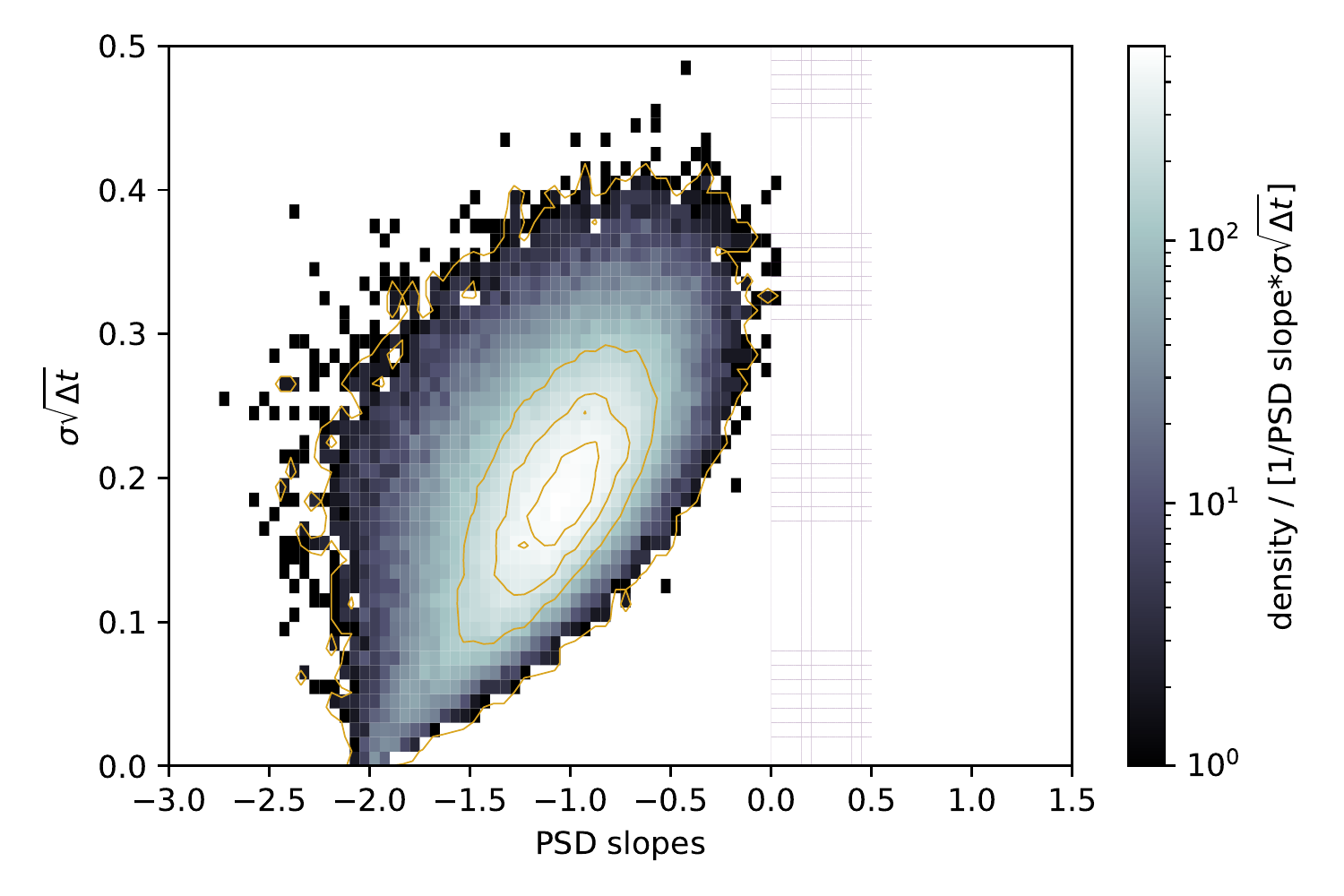}
    \includegraphics[width=0.45\hsize]{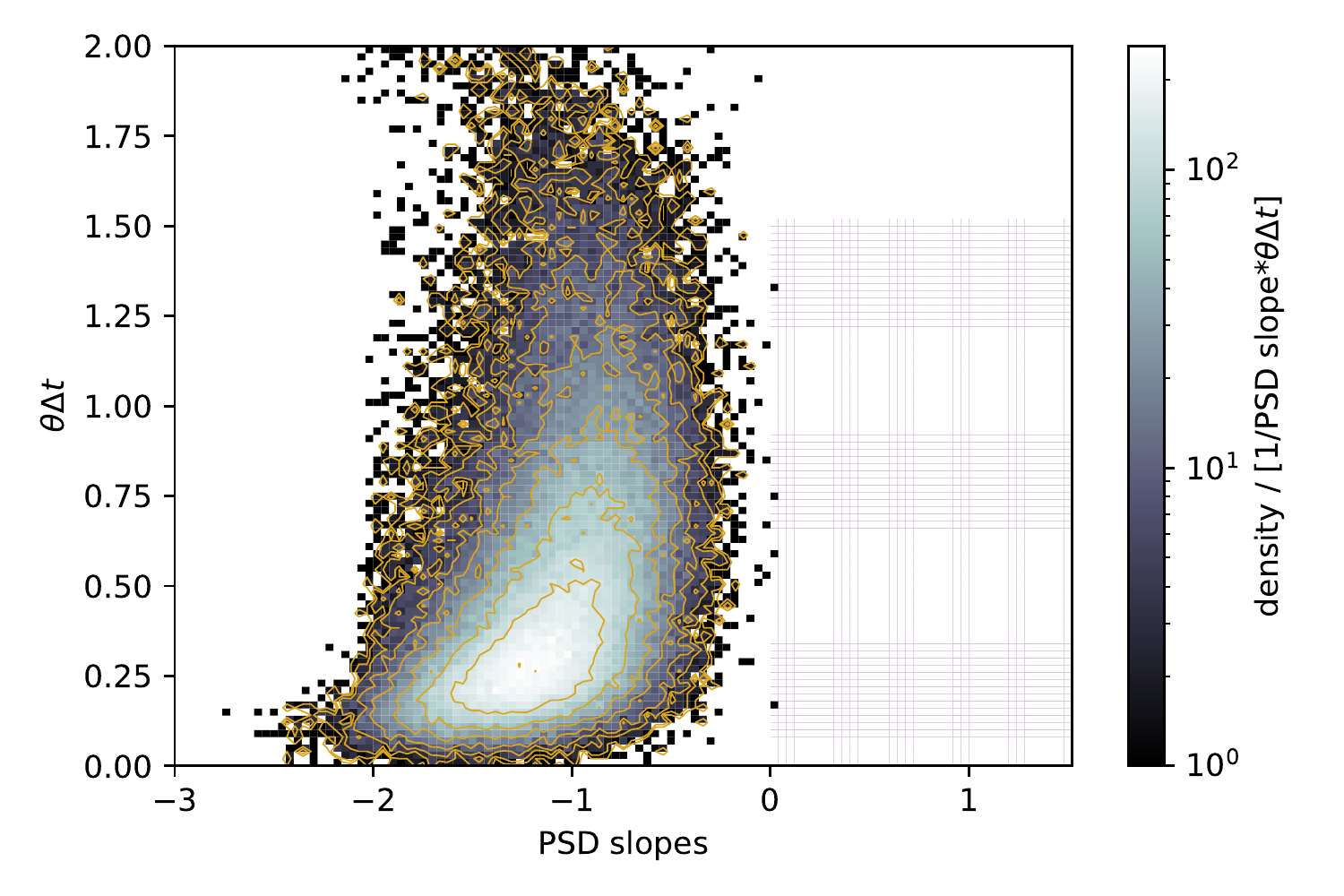}
    \includegraphics[width=0.45\hsize]{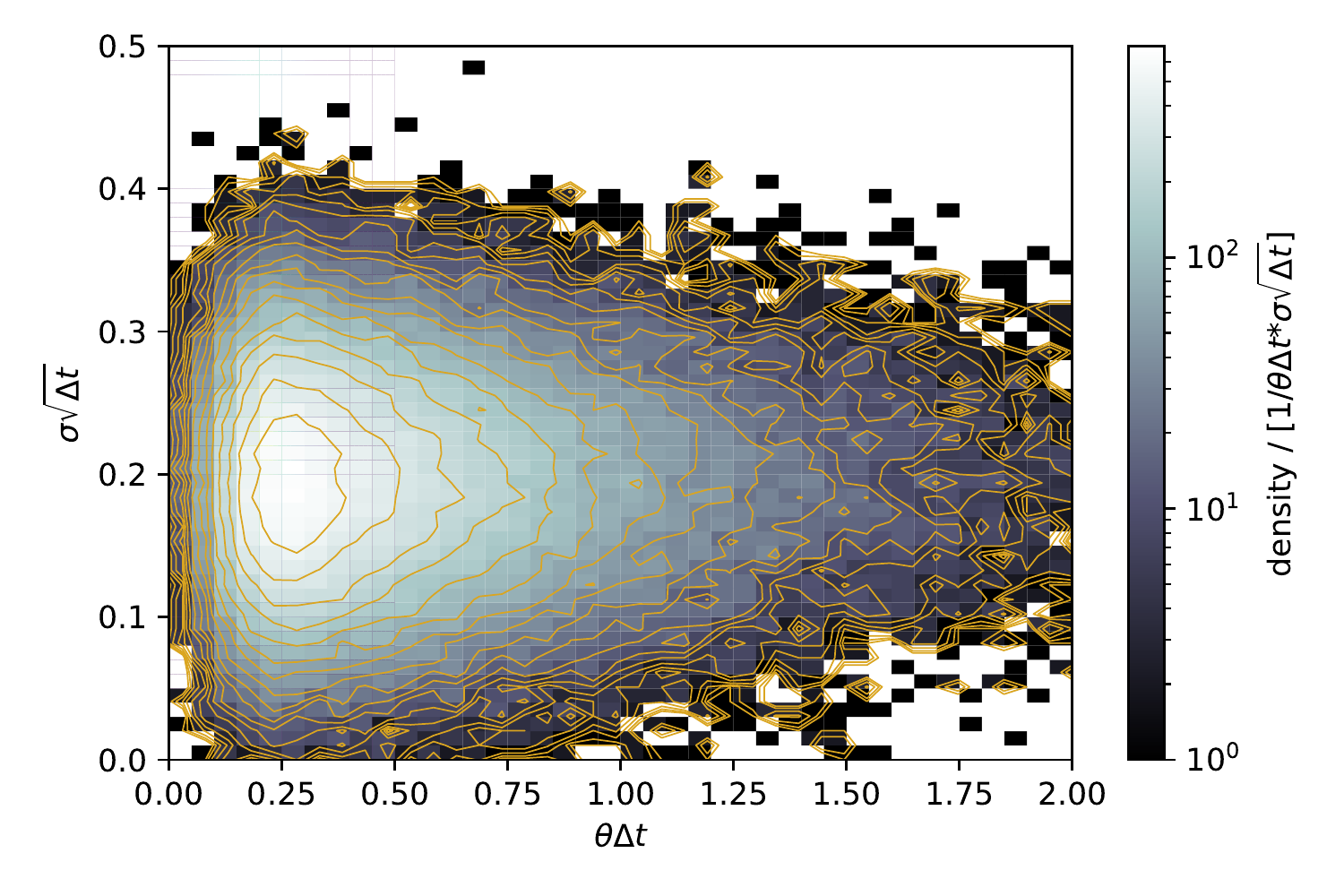}
    \caption{The 2D histograms  illustrate the impact of the OU parameters $\sigma$ and $\theta$ on the PSD slopes. The parameters are set in the simulation for each LC and are obtained from the number generators, as discussed in Subsec.\ref{Subsec.:Number_generators}. The density plots are over plotted with contours, spanning the domain of the color scales and are all separated by a factor of $\sqrt{2}$.  \textit{Top left}: $\sigma$ determines the impact of the white noise term in Eq. (\ref{Eq.Discrete_SDE}). The larger the parameter, the flatter becomes the slope of the PSD. This relation however is not linear (which would be seen as a narrow ellipse) because if $\theta$ is small at the same time when $\sigma$ is large, the white noise terms counteract the effect of $\theta$. Therefore the distribution has a bulky shape. The contour lanes emphasize the tendency of this direct relation between $\sigma$ and the PSD slope , as seen in the ellipses where the data density is the highest. The contour lanes cover a region of $[1,1000]/[1/\text{PSD slope}*\theta \Delta t]$. In the region where the distribution becomes bulky $-2.5 \lesssim \text{PSD slope} \lesssim -1.5$ the effect of $\sigma$ vs. $\theta$ counteraction can be seen. \textit{Top right}: The value of $\theta$ also impacts the PSD slope directly. For $\theta \rightarrow 1$ this however breaks since in this case the only terms left in Eq. (\ref{Eq.Discrete_SDE}) are the white noise terms, where the $\sigma$ parameters takes over. Also in this case the contour lines suggest a direct proportionality between $\theta$ and PSD slopes for $\theta < 0.6$. The contours cover a range of $[1,1000]/ [1/\text{PSD slope}*\theta \Delta t]$. \textit{Bottom}: The relation between $\theta$ and $\sigma$ shows that these values are uncorrelated, however there is a asymmetry in the distribution (positive skewness) where $\theta \rightarrow 1$ where in Eq. (\ref{Eq.Discrete_SDE}) only the white noise terms are left. This skewness explains the crowding of the distributions above. The contours cover a range of $[1,800]/[\theta \Delta t * \sigma \sqrt{\Delta t}]$. }
    \label{Fig:PSD_slopes_theta_sigma}
\end{figure*}

To calculate the PSDs of the observed \Fermi-LCs and the synthetic OU LCs, the Lomb-Scargle algorithm \citep[e.g.][]{Lomb1976,scargle1982,Townsend2010} implemented in SciPy \citep{scipy} is used. Each of the acquired periodograms is fitted with a single power law to obtain the PSD slope $\xi$. Note that this method is insensitive for possible breaks in the PSD slope.  The priority here is to establish the method of OU-parameter extraction for the sample of monthly-binned blazar light curves which can be fitted well with single PSD slopes. It is straightforward to extend the method to more complex PSD shapes produced by mechanisms operating on different spatial scales in the jets.
The $\xi$ distributions of the calibrated OU LCs and the \Fermi-LAT LCs show similar mean values, peaking at $\sim -1$, meaning that most LCs account for power spectra of pink noise while the test LCs peak at $-0.6$. Also the variance, the skew and kurtosis cannot be distinguished. The distributions span a range between $-2\lesssim \xi \lesssim 0$. A pink noise LC typically features fluctuations on short time scales (on the order of several units of time sampling), see \citep[e.g]{timmer_koenig1995,milotti2002,milotti2005}. The $\xi$ -distribution of the test OU LCs show a peak around $0$, while also spanning to $-2$. This results in a variance and skew which are larger by a factor of $\sim 2-3$ compared to the others. This shows that the range of OU parameters must be narrowed down to the range determined with the extraction method to obtain valid synthetic LCs that resemble the observed ones.\\
For the synthetic OU LCs, the impact of the OU parameters on the PSD slope is studied by investigating the density plots of the set OU parameters, obtained for each OU LC as described in Subsec.\ref{Subsec.:Number_generators} and the PSD slopes. Figure \ref{Fig:PSD_slopes_theta_sigma} illustrates the impact of the parameters $\sigma$ and $\theta$ on the PSD slopes. The top left plot shows the relation between $\sigma$ and $\xi$. The larger the impact of the white noise (given larger values of $\sigma$) in Eq. (\ref{Eq.Discrete_SDE}), the flatter the slope. The yellow contour lines in the area of the densest data population indicates this trend, however if $\sigma$ is large and $\theta$ is small, the white noise part and the reversion rate parts counteract each other in Eq. (\ref{Eq.Discrete_SDE}) which can be seen in the distribution becoming bulky at $-2.5 \lesssim \xi \lesssim -0.5$. The plot in the top right shows the impact of $\theta$ on $\xi$. The contour lines might indicate a relation between the reversion rate and PSD slope. The smaller the reversion rate, the steeper the slope. However, this breaks when $\theta \rightarrow 1$. If $\theta = 1$, the reversion rate terms in Eq. (\ref{Eq.Discrete_SDE}) cancel and only the white noise terms remain. The plot in the bottom of Fig. \ref{Fig:PSD_slopes_theta_sigma} suggests that $\sigma$ and $\theta$ are uncorrelated. However, the distribution has a skewness where $\theta \rightarrow 1$. This skewness can explain the crowding in certain areas of the distributions in the plots above.

\begin{figure}
\centering
    \includegraphics[width=\hsize]{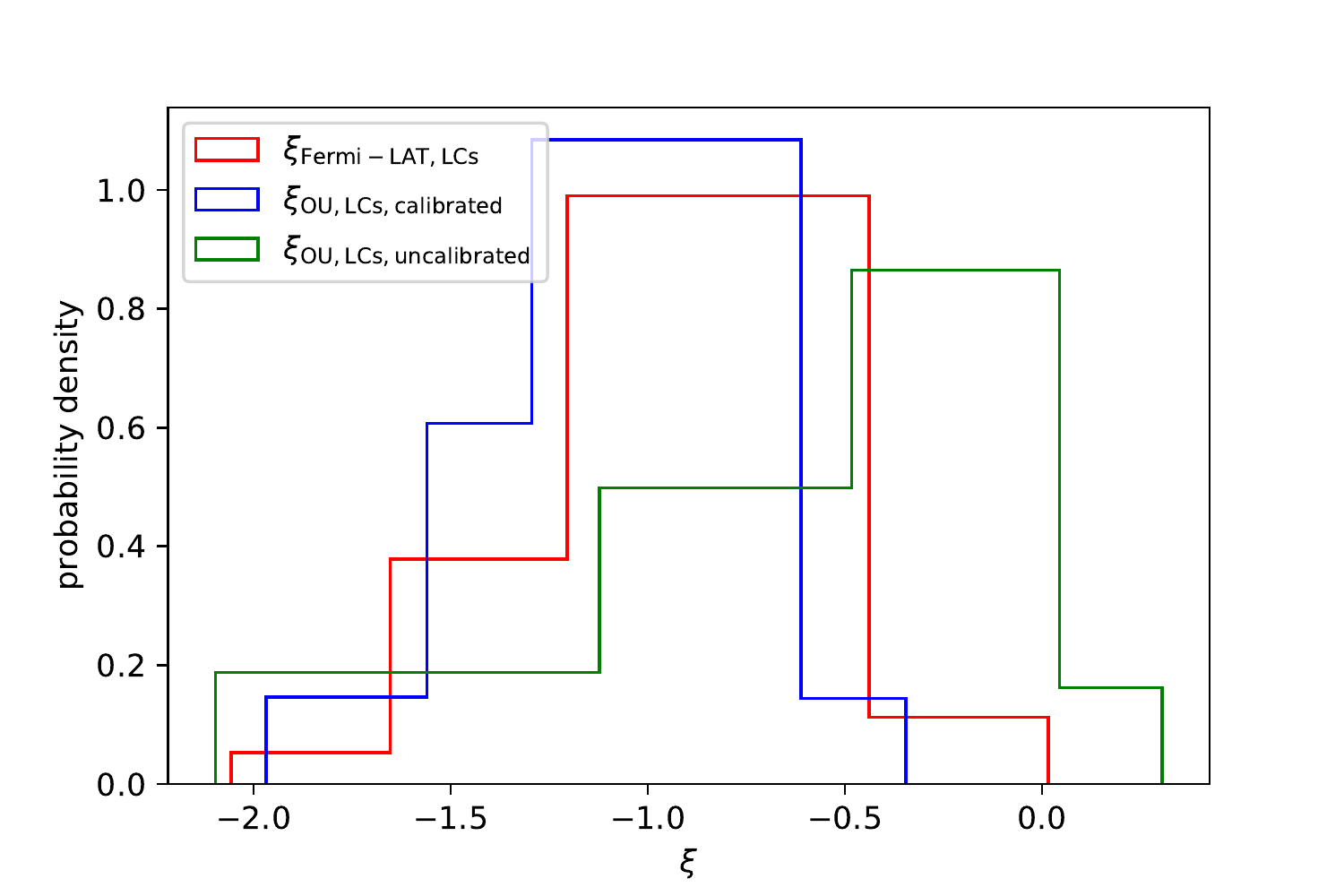}
    \caption{PSD slope distribution for different data sets. The green histogram represents the $\xi$ distribution for the test OU light curves (parameters are drawn from Gaussian distributions) used to test of the extraction method. The red histogram represents the $\xi$ distribution of the observed \Fermi-LAT data. The blue $\xi$ distribution represents the OU LCs where the OU parameters are calibrated with the observed data. The bins are determined with the Bayesian block algorithm \citep{Scargle2013} }
    \label{Fig:PSD_slopes}
\end{figure}


\begin{table}
\caption{Statistical moments of the $\xi$ distributions derived from the observed \Fermi-LCs, the generic test OU LCs as well as the synthetically generated OU LCs based on the extraction of the observed data. The colors of the data sets indicate the colors in Fig.\ref{Fig:PSD_slopes}. col1: data set; col2: mean of the $\xi$ distribution; col3: variance; col4: skew; col5: kurtosis}
\label{tab.:PSD_stats}
\begin{tabular}{lcccc}
\toprule
Data set& Mean& Variance& Skew &Kurtosis\\
\midrule
 \textcolor{red}{\Fermi-LAT} &-0.9& 0.1 & -0.25 & 0.12\\ \midrule
\textcolor{blue}{OU calibrated} &-1.1& 0.09 & -0.26 & 0.09\\ \midrule
\textcolor{green}{OU uncalibrated} &-0.6 & 0.3 & -0.78 & 0.03\\ \bottomrule
 \bottomrule
\end{tabular}
\end{table}

 \subsection{Time asymmetry of flux variations}
 The flux variability of the observed LCs as well as of the synthetic ich OU LCs is further studied by applying the Bayesian block algorithm \citep{Scargle2013}. It determines the best step-wise constant function to describe the data. Based on this, the HOP algorithm \citep{HOP1998} can be used to determine flares with so-called HOP groups, as introduced in \citet{Meyer2019}. It identifies the local maxima of the blocks (peaks) and proceeds downwards as long as the adjacent blocks are subsequently lower analogous to the watershed method of topological data analysis. The onset and offset of a flare is given by the flux exceeding and going under a certain limit. 
 This is motivated by the fact that blazar variability often appears superimposed on a constant (or relatively slowly varying) flux level. For studying intrinsic source variability this background should be removed -- especially if the two components arise from different mechanisms. Noting that the background contributes randomness only through observational errors, e.g. photon counting statistics, assumed to be symmetrically distributed about the true level, \citet{Meyer2019} introduced a robust procedure to determine this ``quiescent background''. The mean flux, since it includes contributions from both, always overestimates the true background. For the sake of this work, we neglect this difference and consider the mean flux of the LC to be the limit for the onset and offset of flares, leaving a more principled approach to future work.
 Each HOP group defines a flare with starting, end, and peak time where the latter is assumed at the center of the highest block in the HOP group. This results in a well defined rise time $t_r$ and decay time $t_d$ for each flare. Based on these, the time asymmetry measure A of the flux variation is given by 
 \begin{equation}
     A = \frac{t_r - t_d}{t_r+t_d}
 \end{equation}
 Three main cases can be distinguished. If $A=0$ there is no difference between rise and decay time as seen for example in a purely Gaussian shot. In the case where $A<0$ the decay time is larger than the rise time. Such ``fast rise - exponential decay'' flare shapes can be seen in gamma-ray bursts \citep[e.g.,][]{Peng2010}. In the case where $A>0$, the rise time is larger than the decay time. 
 This asymmetry measure is derived for each HOP group within the observed 236 LCs. As mentioned above, the mean represents a rather conservative estimate for the limit of a flare. Since the ``quiescent background'' by \citet{Meyer2019} is consistently lower than the mean, additional information about lower flux variations could be obtained by using this as a limit in the HOP algorithm. In future work this will be elaborated on more closely. 
In order to test whether the synthetic LCs generated with the OU based on the parameters extracted from the observed LCs can mimic the latter, we conduct the same procedure 1000 times for 236 synthetic LCs randomly drawn form the entire sample. \\
 HOP groups containing only one Bayesian block are filtered because they yield $A=0$ by definition. Two block HOP groups are conservatively filtered because individual blocks in the observed LCs often contain just few data points.
 Successively, HOP groups containing, three, four or less Bayesian blocks are omitted to study the systematic behind the method and compare synthetic and observed asymmetry measures via a KS test. For each set the fraction of p-values indicating the $5\%$ $(2\,\sigma)$  and the $0.3\%$ ($3\, \sigma$) threshold, possibly inferring difference between the synthetic and observed asymmetry measures, is determined. The results are listed in Tab.\ref{tab.:p-val_fractions}. The fractions of LCs where the asymmetry measures differ at a $2\sigma$ level is ($\sim 37\%$,$\sim 12\%$) for three and four block HOP omitted. At the $2\sigma$ level the fractions are ($\sim 4\%$,$\sim 0.6\%$). Overall the asymmetry measure for the synthetic OU and the observed \Fermi-LCs match well. 
\begin{figure}
\centering
        \includegraphics[width=0.45\hsize]{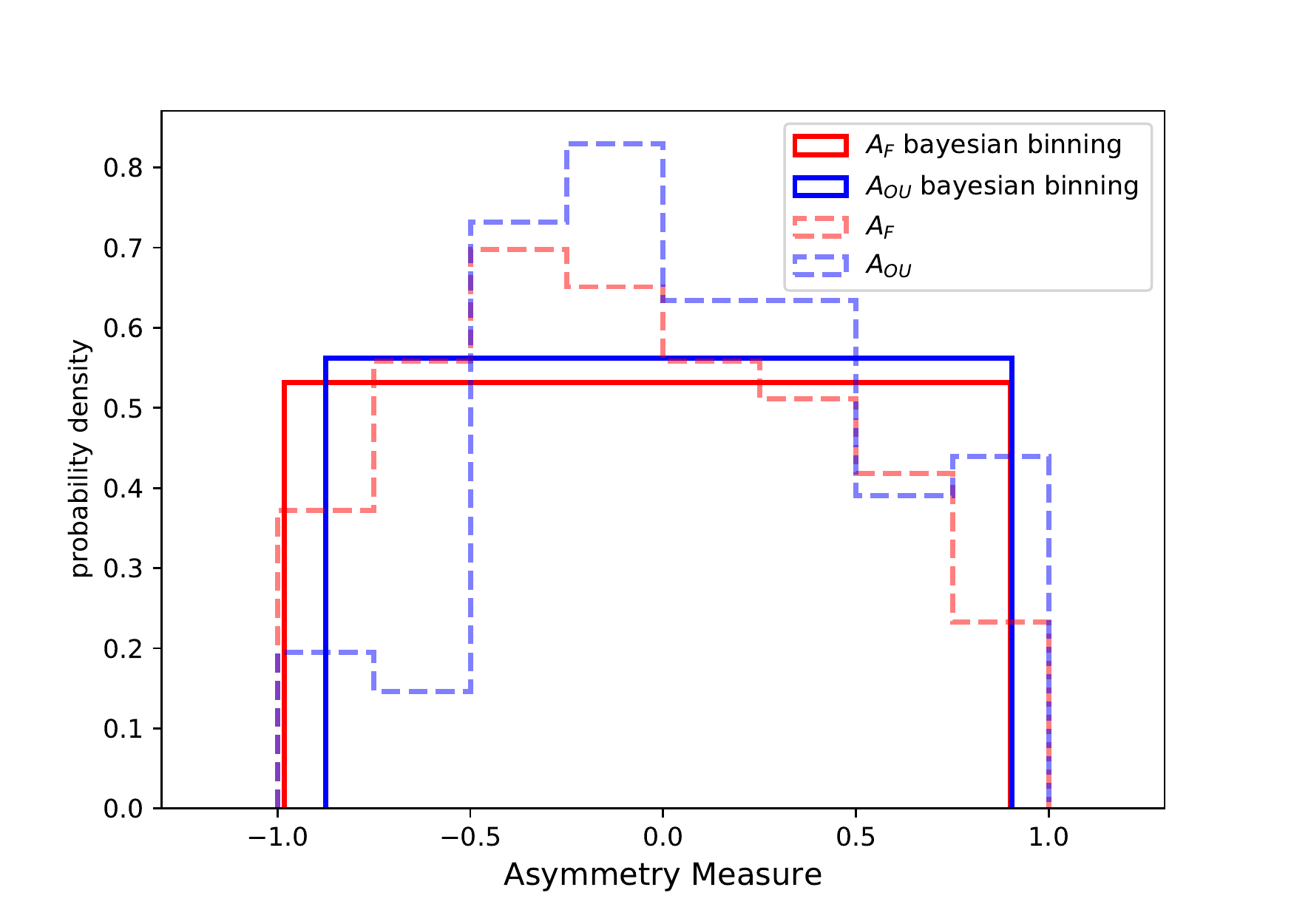}
        \includegraphics[width=0.45\hsize]{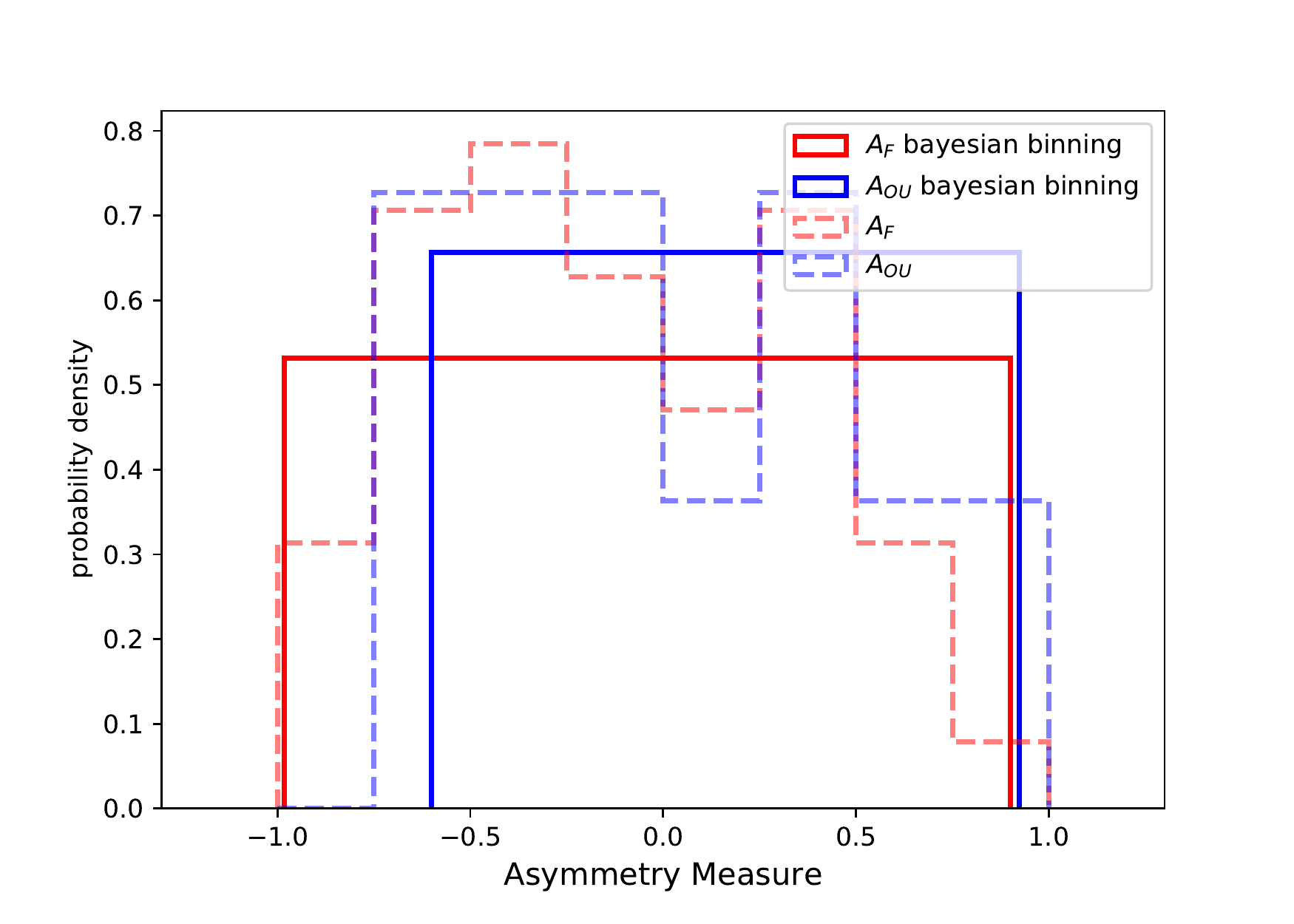}

    \caption{Asymmetry measures for the observed LAT LCs compared to a subset of synthetic OU LCs. The bins are set by the Bayesian block algorithm. The dashed and transparent bins are set to 0.25 and are plotted for convenience to show a 'classical' histogram. The Bayesian binning shows that the asymmetry measures for both data sets basically are present in a range between (-1,1) with the same probability. Top: single block HOP groups omitted; Mid: two-block HOP groups omitted; Bottom: three-block HOP groups omitted. }
    \label{Fig:rise_dacay}
\end{figure}

 \begin{table}
\caption{The fraction of the data sets where a statistical difference is hinted in the asymmetry measure is listed for flares containing more than one two or three blocks .}
\label{tab.:p-val_fractions}
\begin{tabular}{lcc}
\toprule
Data set& $2\, \sigma$ & $3\, \sigma$\\
\midrule
three block flares omitted  & $36.7\%$ & $3.7\%$ \\ \midrule
four block flares omitted & $11.6\%$ & $0.6\%$ \\  \bottomrule

\end{tabular}
\end{table}

\subsection{Examples}
\begin{figure*}
\centering
        \includegraphics[width=0.45\hsize]{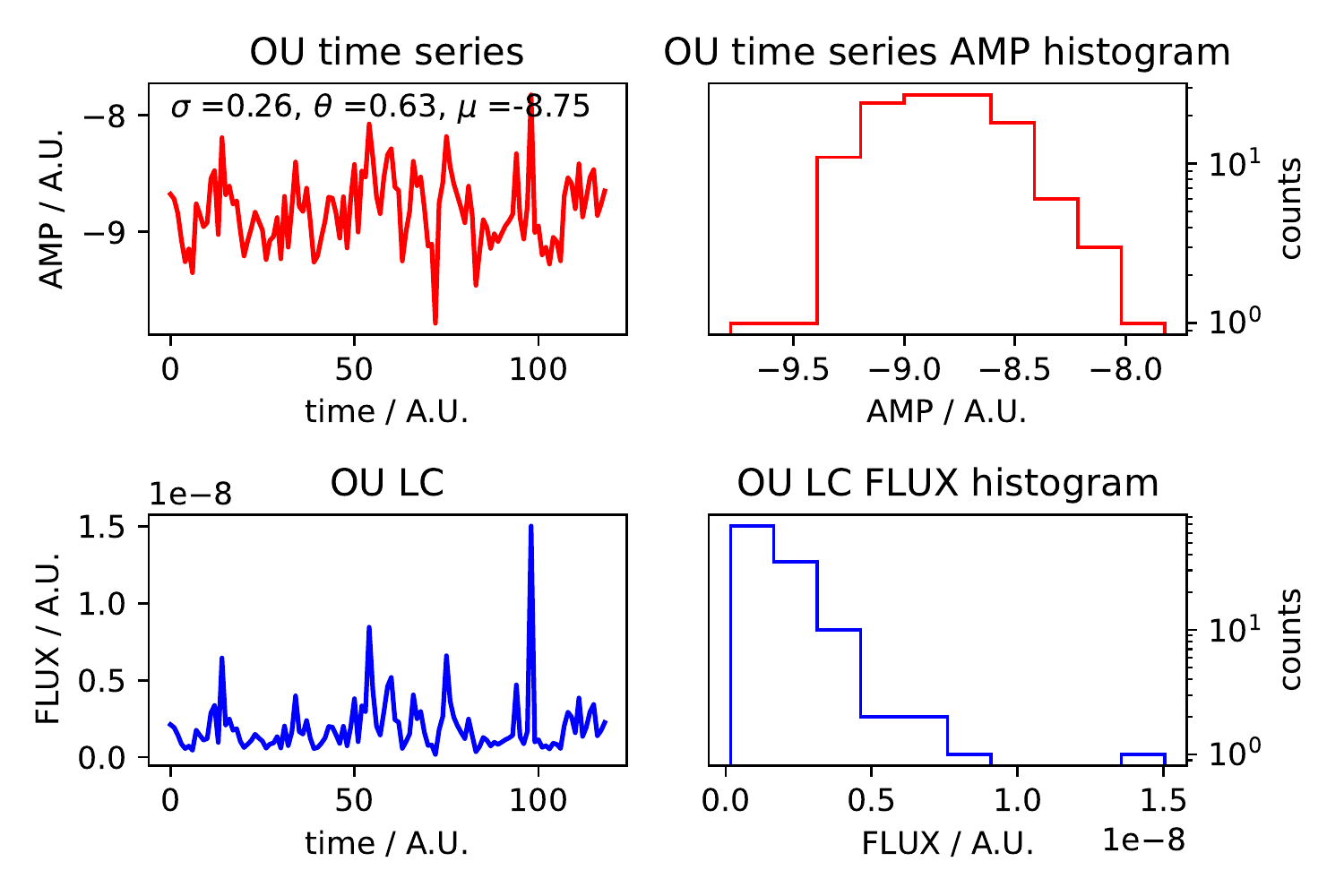}
        \includegraphics[width=0.45\hsize]{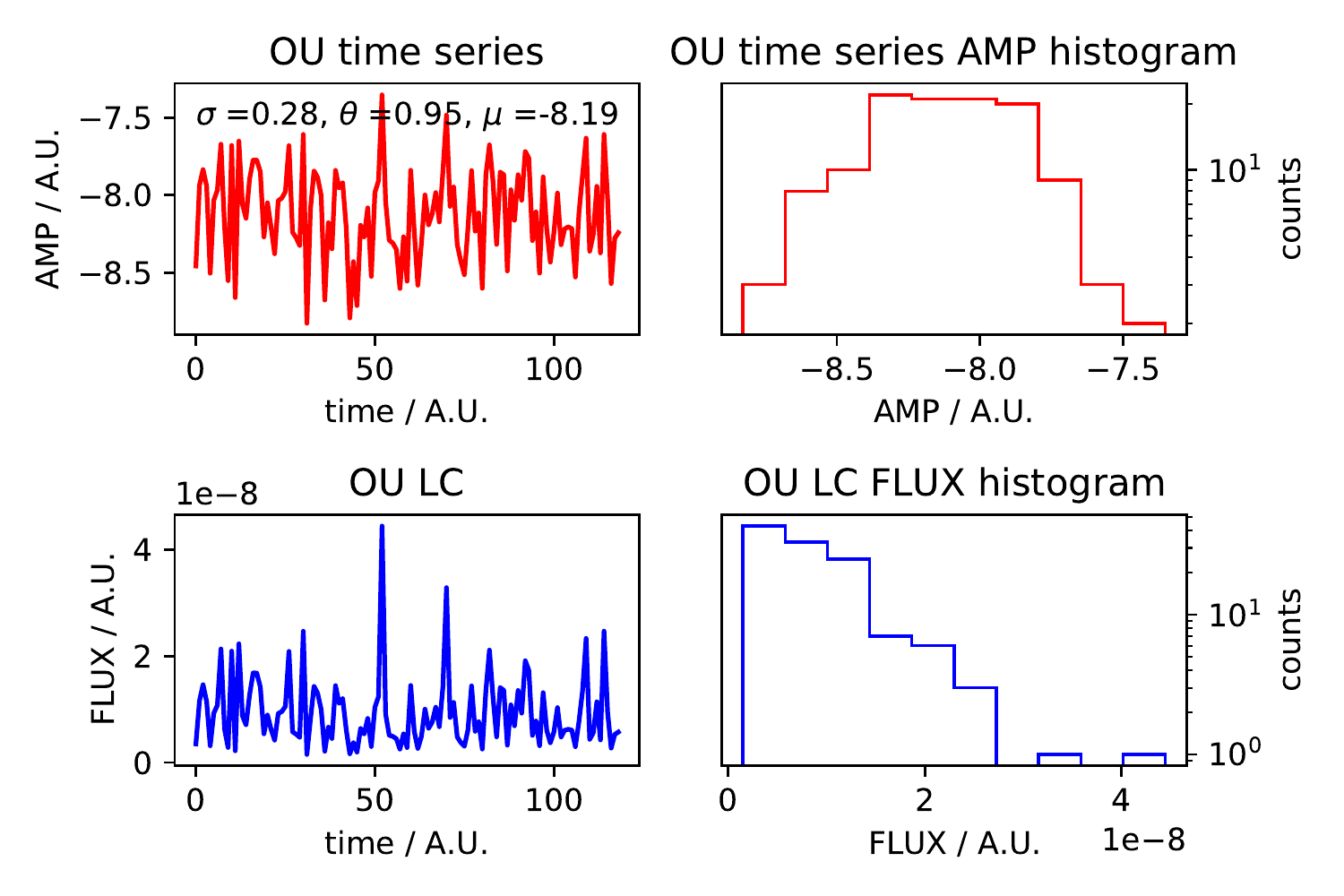}
        \includegraphics[width=0.45\hsize]{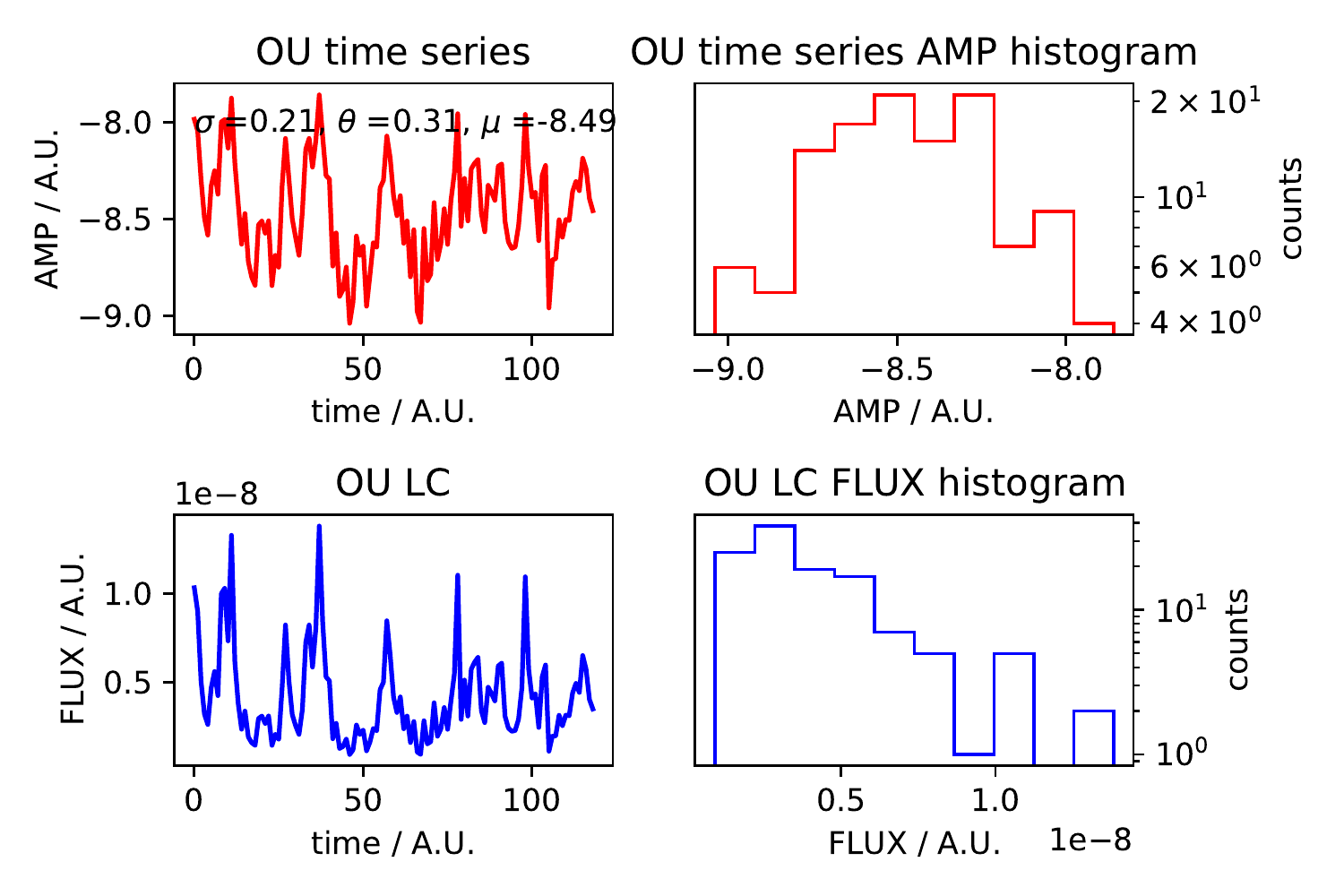}
        \includegraphics[width=0.45\hsize]{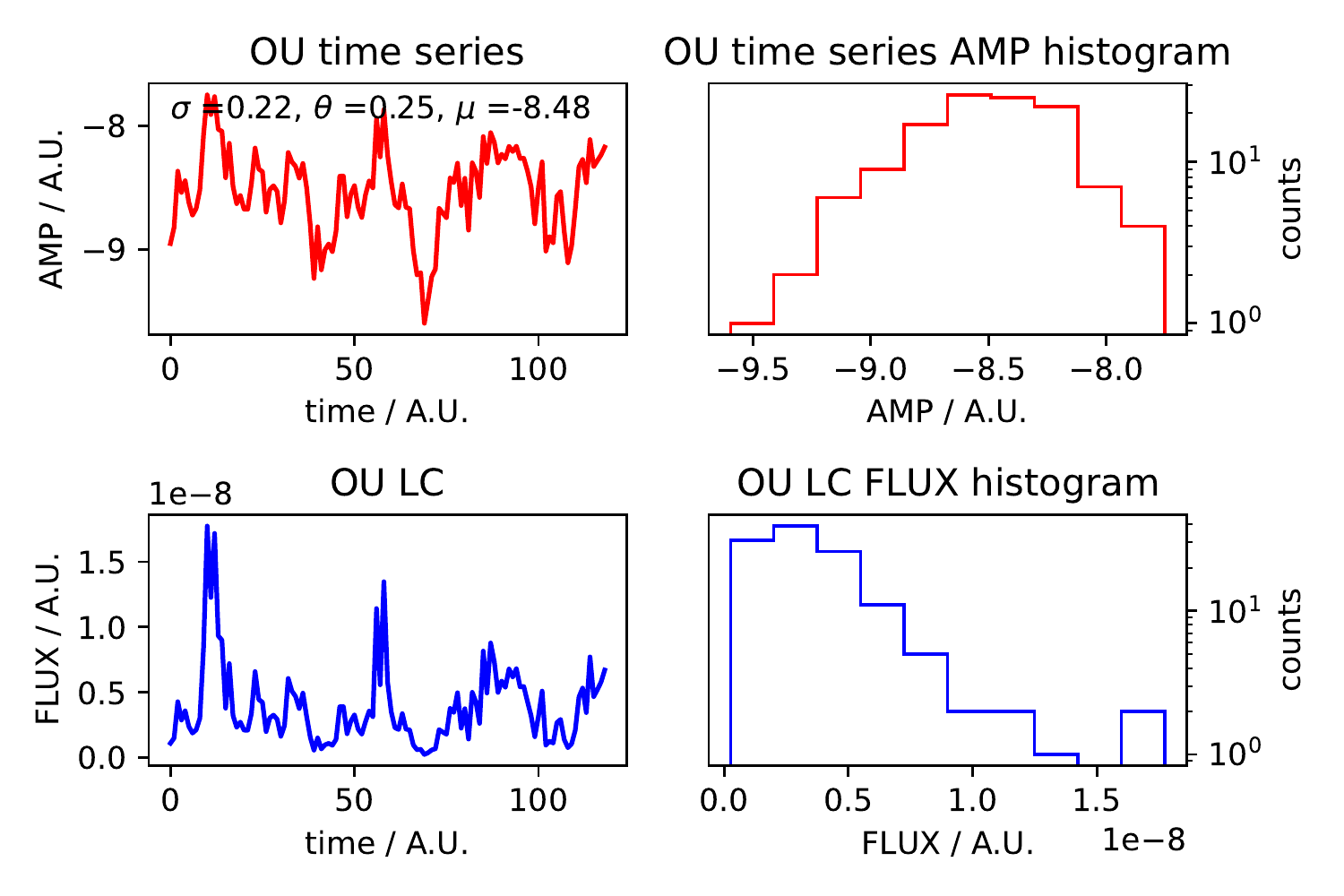}
        \includegraphics[width=0.45\hsize]{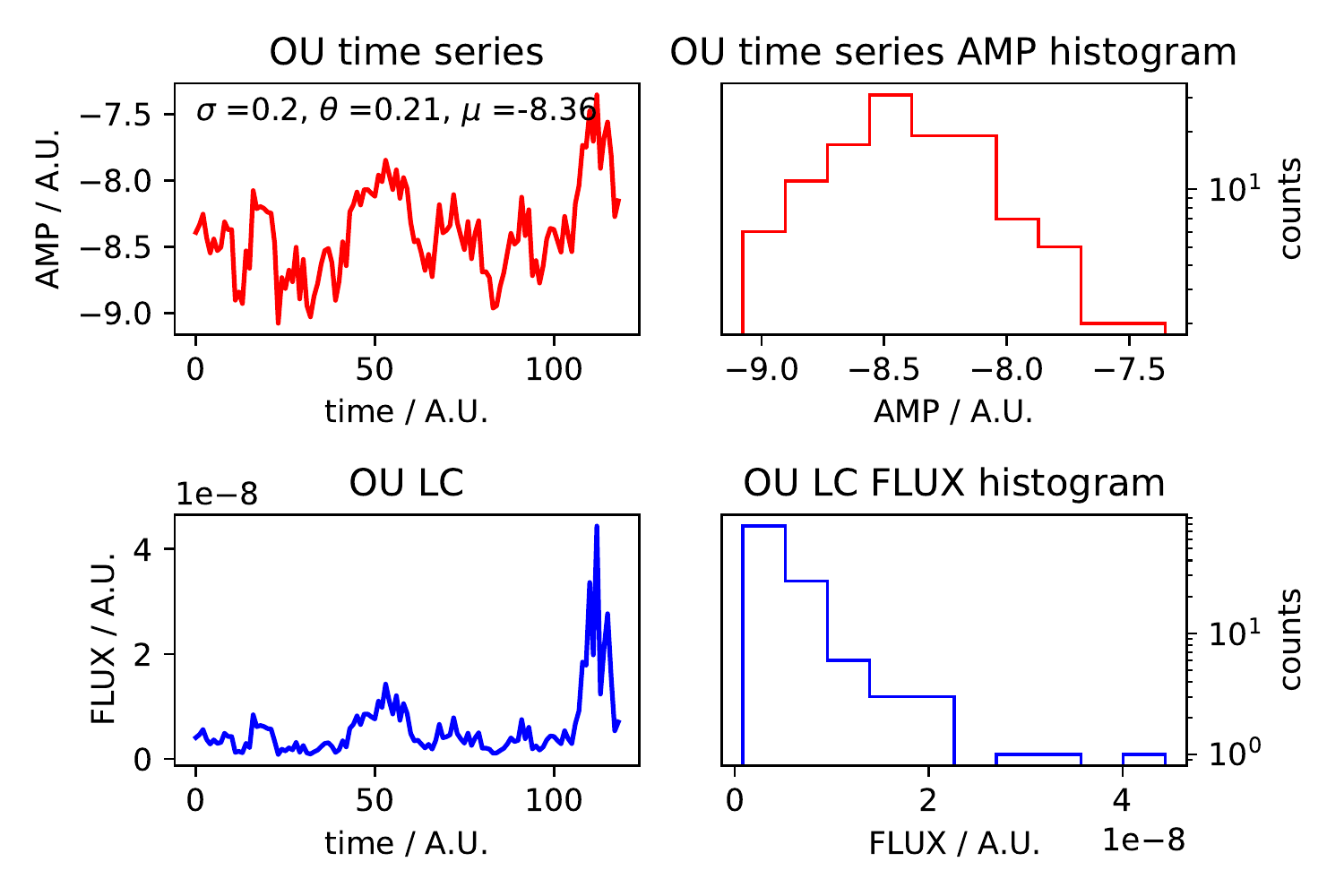}
        \includegraphics[width=0.45\hsize]{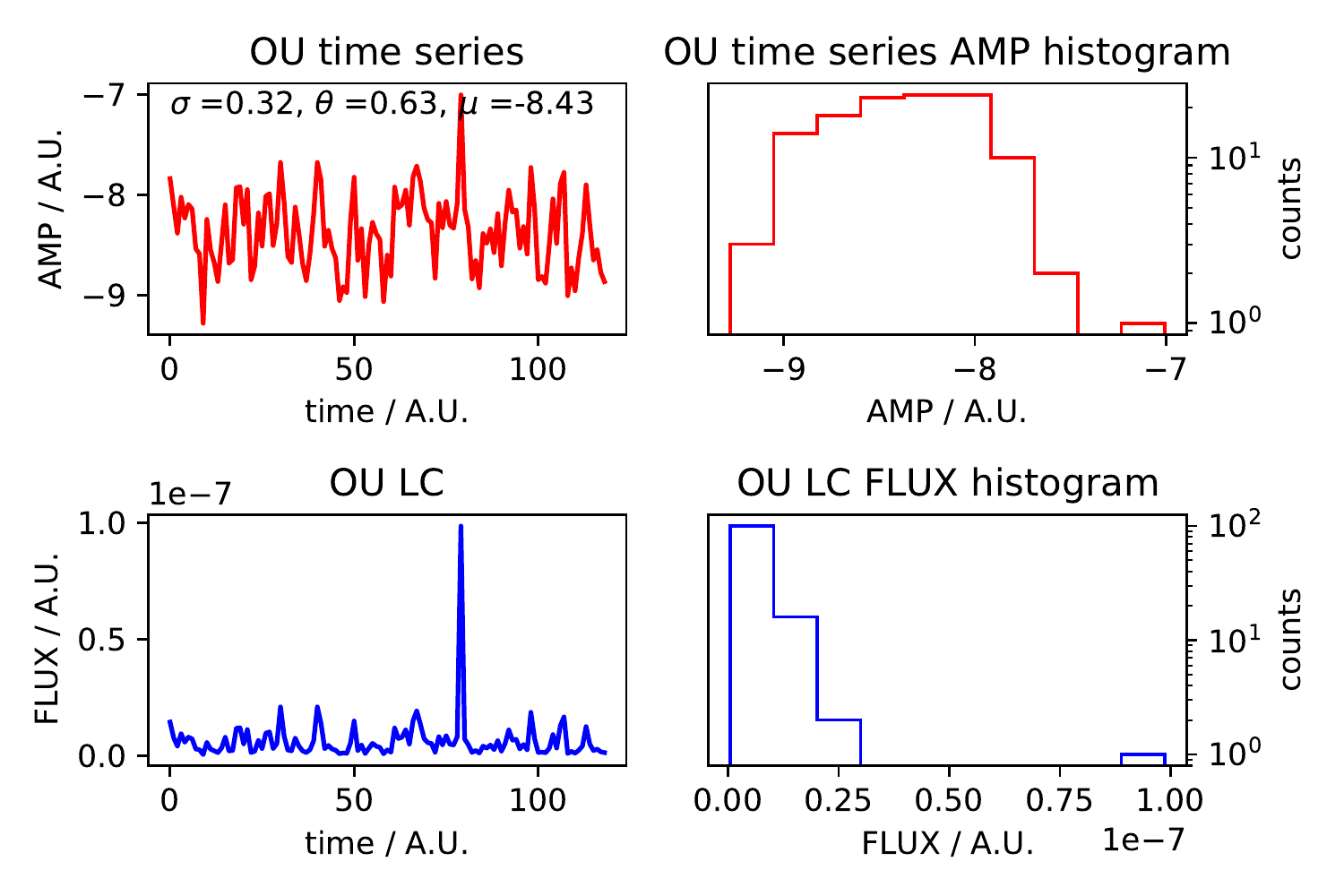}

    \caption{Representative examples of LCs generated with the extracted OU parameters are shown. Each block illustrates a time series and its histogram (red) and the corresponding LC with its histogram (blue). The OU parameters used to simulate the corresponding data sets are shown in the top right.
    The comparison of these plots show that the OU process allows to re-create LCs, that show various kinds of colored noise ($-2\leq  (\xi) \leq 0$, red-pink-white). \textit{Top left}: $\xi \sim -0.2$; \textit{Top right}: $\xi \sim -0.5$; \textit{mid left}: $\xi \sim -1$; \textit{mid right}: $\xi \sim -1.5$; \textit{bottom left}: $\xi \sim -2$; \textit{bottom right}: $\xi \sim -0.2$}
    \label{Fig:examples}
\end{figure*}

To show that the OU process is capable of reproducing the full range of properties in the observed LCs, Fig.\ref{Fig:examples} illustrates sources with different PSD slopes. The OU LCs with pink-ish noise feature flickering noise on short time scales, while the LCs with steeper PSD slopes tend to show variability on large time scales (compared to the binning time scale). As the PSD slope becomes steeper, the variability changes from shorter (flickering) time scales (pink) noise to variability on larger time scales (red) noise. The LC in the bottom left is a peculiar case of a white noise LC featuring one major outbreak within the simulation (observed) time span. 

\section{Conclusion and outlook}
We have shown that the statistical properties featured in \Fermi-LAT LCs can be reproduced with a stationary (exponential) OU process. 
The key step to achieving this result was to develop a method for the extraction of the best-fit OU parameters from the comparison of simulated and observed LCs. In detail our findings are:

\begin{itemize}
    \item Fitting the observed monthly-binned \Fermi-LAT LCs with a stationary (exponential) OU process yields the parameters of equivalent synthetic LCs.
    \item The best-fit set of OU parameters is given by $(\sigma, \mu,\theta)=(0.2,-8.4,0.5)$.
    \item The flares extracted from the observed \Fermi-LAT LCs do not show any preference for shorter or longer rise time scales compared to their decay time scales which is also reproduced by the OU LCs. This is  unexpected if the rise times are determined by an acceleration process and the decay times by cooling which in general are very different unless not  exactly balanced at the observed energy.
    \item The OU process can reproduce the range of colored noise (PSD slopes)  of \Fermi-LAT LCs, including the LCs of sources showing flickering variability, variability on larger time scales, and even sources that show only one large outbreak within the simulation (observation) time span.
    \item We note that the OU process describes the random fluctuations of a logarithmic energy flux which indicates that the underlying physical mechanism amplifies random fluctuations exponentially. Such processes have been discussed in models of the non-linear evolution of fluid-dynamical instabilities leading to particle acceleration by magnetic reconnection (\cite{2016ApJ...828...92Y,2017SSRv..207..291B,2019MNRAS.482...65C}) or the runaway acceleration of particles confined between random sequences of colliding shock waves triggered by fluctuations of the jet velocity \citep{2019ApJ...887..133B}.
\end{itemize}

Our results have been obtained using monthly-binned \Fermi-LAT LCs $E>1\, \mathrm{GeV}$ of 236 blazars, which cover a time interval of approximately ten years.  The binning time scale exceeds the minimum variability time scales of blazars, which can be as short as minutes, by a large margin. The LCs apparently track secular changes of the energy dissipation on the dynamical scale of sub-parsec jets associated with fluid-dynamical instabilities and non-thermal flux amplification due to particle acceleration events.  Further diagnosis, using shorter time-scale binning and multi-frequency monitoring data will be needed to identify the physical origin of the observed randomness and its limitations by searching for short-range correlations in the time- and frequency domains. The methods described in this paper and their implementation are versatile tools to carry out such studies.

\section{Appendix A}
\label{Sec.:AppendixA}
\begin{subequations}
\subsection{Proof of validity of explicit formula for the discrete SDE}
\label{AppA1}
By definition the initial case is fulfilled with.
    \begin{equation}
    x(0) = x(0) 
    \end{equation}
    Thus the induction can be initialized
    \begin{equation}
    x(s+1) = (1-\theta \Delta t)x(s) + \theta \Delta t \mu + \sigma \sqrt{\Delta t} N(0, 1, \Delta t s),
    \end{equation}
    \begin{equation}
    \begin{gathered}
    x(s+1) = (1-\theta\Delta t) \Bigg{[} (1-\theta \Delta t)^{s}x(0) + \\ \sum _{k=1}^{s}\Big((1-\theta\Delta t)^{s-k}(\theta \Delta t \mu + \sigma \sqrt{\Delta t} N(0, 1, (k-1)\Delta t)\Big) \Bigg{]}\\
    + \theta \Delta t \mu + \sigma \sqrt{\Delta t} N(0, 1, \Delta t s),
    \end{gathered}
    \end{equation}
    \begin{equation}
    \begin{gathered}
    x(s+1) = (1-\theta \Delta t)^{s+1}x(0) + (1-\theta \Delta t)\\
    \sum_{k=1}^{s}\left((1-\theta\Delta t)^{s-k}(\theta \Delta t \mu + \sigma \sqrt{\Delta t} N(0, 1, (k-1)\Delta t)\right)\\
    + \theta \Delta t \mu + \sigma \sqrt{\Delta t} N(0, 1, \Delta t s),
    \end{gathered}
    \end{equation}
    \begin{equation}
    \begin{gathered}
    x(s+1) = (1-\theta \Delta t)^{s+1}x(0)+ \\
    \sum_{k=1}^{s}\left((1-\theta\Delta t)^{s+1-k}(\theta \Delta t \mu + \sigma \sqrt{\Delta t} N(0, 1, (k-1)\Delta t)\right) \\
    + \theta \Delta t \mu + \sigma \sqrt{\Delta t} N(0, 1, \Delta t s),
    \end{gathered}
    \end{equation}
    \begin{equation}
    \begin{gathered}
    s(x+1) = (1-\theta \Delta t)^{s+1}x(0)+\\ \sum_{k=1}^{s+1}\left((1-\theta\Delta t)^{s+1-k}(\theta \Delta t \mu + \sigma \sqrt{\Delta t} N(0, 1, (k-1)\Delta t)\right) \square.
    \end{gathered}
    \end{equation}
\end{subequations}

\subsection{Proof of the 1st criterion for stationarity}
\label{AppA2}
To prove the first criterion (Eq. \ref{Eq.Stat1}), Eq. (\ref{Eq.proof-stat-mean-ansatz}) is shown to hold for all $s$, where $s$ is an positve integer
\begin{subequations}
    \begin{equation}
    \langle x(s)\rangle \overset{!}{=} \langle x(s+1)\rangle. 
    \label{Eq.proof-stat-mean-ansatz}
    \end{equation}
Hence $x(s)$ can be written as normal distributions (Eq. \ref{Eq.x-normal}), the mean is the first argument of $N$. Therefore Eq. (\ref{Eq.proof-stat-mean-ansatz}) simplifies to. 
    \begin{equation}
	\begin{gathered}
	(1-\theta \Delta t)^{s} x(0) + \theta \Delta t \mu \sum_{n=0}^{s-1}{(1-\theta \Delta t)^{n}} =\\ 
	(1-\theta \Delta t)^{s+1} x(0) + \theta \Delta t \mu \sum_{n=0}^{s}{(1-\theta \Delta t)^{n}},
	\end{gathered}
    \end{equation}
    \begin{equation}
    0 = [(1-\theta \Delta t)^{s+1}-(1-\theta \Delta t)^{s}] x(0) + \theta \Delta t \mu (1-\theta \Delta t)^{s},
    \end{equation}
    \begin{equation}
    \underbrace{(1-\theta \Delta t)^{s+1} x(0)}_{\mathrm{LHS}} = \underbrace{(x(0) -\theta \Delta t \mu)(1-\theta \Delta t)^{s}}_{\mathrm{RHS}}.
    \end{equation}
Let
    \begin{equation}
    \begin{gathered}
    x(0) = \mu \Rightarrow \mathrm{RHS} = (\mu -\theta \Delta t \mu) (1-\theta\Delta t)^s = \mu (1-\theta\Delta t)^{s+1} =\\ \mathrm{LHS}|_{x(0)=\mu} \square.
    \label{Eq.proof-stat-mean-result}
    \end{gathered}
    \end{equation}

If a system runs sufficiently long, the criterion of \ref{Eq.proof-stat-mean-result} loses on importance, because after a sufficient amount of time a steady-state is reached. Therefore the system is insensitive to the initial conditions. Hence
    \begin{equation}
    \mu = x(0) = \langle x(0) \rangle = \langle x(s) \rangle.
    \label{Eq.mean}
    \end{equation}
\end{subequations}

\subsection{Proof of the 2nd criterion for stationarity}
\label{AppA3}
For the proof of \ref{Eq.Stat2} the normal distributed nature of$x(s)$ is used and that $\langle d^{2} \rangle = Var(d)$ for a given distribution distribution $d$, 
\begin{subequations}
    \begin{equation}
    \mathrm{Var}(x(s)) = \sigma^{2}\Delta t \sum_{n=0}^{s-1}{(1-\theta \Delta t)^{2n}}.
    \label{Eq.var-tot-1}
    \end{equation}
This expression is required to be finite for all times. Therefore a lower limit of $s=0$ is given and the analysis of the limit $s\rightarrow\infty$ suffices
    \begin{equation}
	\begin{gathered}
    Var(x(s)) =\lim\limits_{s \rightarrow \infty}{\sigma^{2}\Delta t \sum_{n=0}^{s-1}{(1-\theta \Delta t)^{2n}}} 
    =\lim\limits_{s \rightarrow \infty}{\sigma^{2}}\Delta t \\ \sum_{n=0} ^{s-1} {[(1-\theta \Delta t)^{2}]^{n}},
	\end{gathered}
    \end{equation}
    \begin{equation}
	Var(x(s))\overset{\mathrm{geo.\; series}}{=}\begin{cases}
	\frac{\sigma^{2} \Delta t}{1-(1-\theta \Delta t)^{2}}, & (1-\theta \Delta t)^{2} < 1 \\
	\infty, & (1-\theta \Delta t)^{2} \geq 1 \\
	\end{cases},
    \label{Eq.var-tot-2}
    \end{equation}
    \begin{equation}
	Var(x(s))=\begin{cases}
	\frac{\sigma^{2} \Delta t}{2 \theta \Delta t-\theta^{2} \Delta t^{2}}, & |1-\theta \Delta t| < 1 \\
	\infty, & |1-\theta \Delta t| \geq 1 \\
	\end{cases}.
    \label{Eq.stat-3}
    \end{equation}
\end{subequations}
Thus a stationary process fulfills this requirement for $|1-\theta \Delta t| < 1$. Also the parameter space $\theta \Delta t$ can be constrained with this expression.
\subsection{Proof of the 3rd criterion for stationarity}
\label{AppA4}
For the following proof it is necessary to express $x(t+\tau)$, dependent on $x(t)$ and independent form $x(0)$. 
\begin{subequations}
    \begin{equation}
    \begin{gathered}
    x(s+\tau)=(1-\theta \Delta t)^{s+\tau}x(0)+\\ \sum_{k=1}^{s+\tau}{\left[(1-\theta \Delta t)^{s+\tau-k}\cdot (\theta \Delta t \mu+\sigma\sqrt{\Delta t} N(0, 1, (k-1)\Delta t))\right]}
    \end{gathered}
    \end{equation}
Separating the sums, from index $1$ to $s$ and from $s+1$ to $s+\tau$ yields
    \begin{equation}
    \begin{gathered}
     x(s+\tau)= (1-\theta\Delta t)^{s}(1-\theta\Delta t)^{\tau}x(0)+ \\
    \sum_{k=1}^{s}{\left[(1-\theta\Delta t)^{s-k}(1-\theta\Delta t)^{\tau}\cdot (\theta \Delta t \mu+\sigma\sqrt{\Delta t} N(0, 1, (k-1)\Delta t)\right]}+ \\
    \sum_{k=s+1}^{s+\tau}{\left[(1-\theta \Delta t)^{s+\tau-k}\cdot (\theta \Delta t \mu+\sigma\sqrt{\Delta t} N(0, 1, (k-1)\Delta t))\right]}.
    \end{gathered}
    \end{equation}
Factorising $(1-\theta \Delta t)^{\tau}$ from the first two terms yields 
    \begin{equation}
    \begin{gathered}
    x(s+\tau)=(1-\theta\Delta t)^{\tau}\\\underbrace{\left((1-\theta\Delta t)^{s}x(0)+\sum_{k=1}^{s}{\left[(1-\theta\Delta t)^{s-k}(\theta \Delta t \mu+\sigma\sqrt{\Delta t} N(0, 1, (k-1)\Delta t)\right]}\right)}_{x(s)} \\
    +\sum_{k=s+1}^{s+\tau}{\left[(1-\theta \Delta t)^{s+\tau-k}\cdot (\theta \Delta t \mu+\sigma\sqrt{\Delta t} N(0, 1, (k-1)\Delta t))\right]},
    \end{gathered}
    \end{equation}
where $x(s)$ substitutes the under braced part. The indices of the remaining sum are shifted by $s$
    \begin{equation}
    \begin{gathered}
    x(s+\tau) = (1-\theta\Delta t)^{\tau}\cdot x(s)
    +\\ \sum_{k=1}^{\tau}{\left[(1-\theta \Delta t)^{\tau-k}\cdot (\theta \Delta t \mu+\sigma\sqrt{\Delta t} N(0, 1, (k+s-1)\Delta t))\right]}.
    \end{gathered}
    \end{equation}
\end{subequations}
With the explicit expression for $x(s+\tau)$, the thrid criterion (Eq. \ref{Eq.Stat3}) can be given.
\begin{subequations}
    \begin{equation}
    \begin{gathered}
    \langle x(s) x(s+\tau) \rangle =
    \bigg\langle x(s) \bigg( (1-\theta\Delta t)^{\tau} x(s)
    +\\ \sum_{k=1}^{\tau}{\left[(1-\theta \Delta t)^{\tau-k} (\theta \Delta t \mu+\sigma\sqrt{\Delta t} N(0, 1, (k+s-1)\Delta t))\right]} \bigg) \bigg\rangle.
    \end{gathered}
    \end{equation}
Multiplying $x(s)$ in the sum and the linearity of the expectation value yields
    \begin{equation}
    \begin{gathered}
    \langle x(s) x(s+\tau) \rangle = \\
     \left\langle (1-\theta\Delta t)^{\tau} x^{2}(s) \right\rangle + \left\langle \sum_{k=1}^{\tau}{\left[ x(s) (1-\theta\Delta t)^{\tau-k} \theta\Delta t \mu x(s) \right]} \right\rangle + \\ \left\langle \sum_{k=1}^{\tau}{(1-\theta\Delta t)^{\tau-k} \sigma \sqrt{\Delta t} N(0, 1, (k+s-1)\Delta t)} \right\rangle,
    \end{gathered}
    \end{equation}
utilizing the linearity of the expectation value
    \begin{equation}
    \begin{gathered}
    \langle x(s) x(s+\tau) \rangle = \\
     (1-\theta\Delta t)^{\tau}\left\langle x^{2}(s) \right\rangle + \sum_{k=1}^{\tau}{\left[ (1-\theta\Delta t)^{\tau-k} \theta\Delta t \mu \left\langle x(s) \right\rangle \right]}  + \\ \left\langle \sum_{k=1}^{\tau}{x(s) (1-\theta\Delta t)^{\tau-k} \sigma \sqrt{\Delta t} N(0, 1, (k+s-1)\Delta t)} \right\rangle.
     \label{Eq.app14c}
    \end{gathered}
    \end{equation}

\rule{\columnwidth}{.5pt}
Auxiliary to simplify the last summand:
\begin{subequations}
    \begin{equation}
    \left\langle \sum_{k=1}^{\tau}{x(s) (1-\theta\Delta t)^{\tau-k} \sigma \sqrt{\Delta t} N(0, 1, (k+s-1)\Delta t)} \right\rangle
    \label{eq:aux1}
    \end{equation}
Replace $x(s)$ in Eq. (\ref{eq:aux1}) with the explicit form. 
    \begin{equation}
    \begin{gathered}
     \Bigg\langle \sum_{k=1}^{\tau}{\Bigg[\sum_{i=1}^{s}{\left[ (1-\theta\Delta t)^{s}(\theta\Delta t \mu + \sigma\sqrt{\Delta t}N(0, 1, (i-1)\Delta t)) \right]}} \,\cdot \\
    (1-\theta\Delta t)^{\tau-k} \sigma \sqrt{\Delta t} N(0, 1, (k+s-1)\Delta t)\Bigg] \Bigg\rangle.
    \end{gathered}
    \end{equation}
Expending the product inside the sum into the inner sum yields
    \begin{equation}
    \begin{gathered}
     \Bigg\langle \sum_{k=1}^{\tau}{\sum_{i=1}^{s}{\Big[(1-\theta\Delta t)^{s+\tau-k}\theta\Delta t \sigma \sqrt{\Delta t} N(0, 1, (k+s-1)+ }}\\(1-\theta\Delta t)^{s+\tau-k} \sigma^{2} \Delta t N(0, 1, (i-1)\Delta t) N(0, 1, (k+s-1)\Delta t\Big] \Bigg\rangle.
    \end{gathered}
    \end{equation}
Using linearity, the expectation value of each of the inner summands can be calculated separately. 
    \begin{equation}
    \begin{gathered}
     \left\langle \sum_{k=1}^{\tau} \sum_{i=1}^{s} \Big[(1-\theta\Delta t)^{s+\tau-k}\theta\Delta t \sigma \sqrt{\Delta t} N(0, 1, (k+s-1)\Big] \right\rangle+ \\ \Bigg\langle \sum_{k=1}^{\tau}\sum_{i=1}^{s}\Big[(1-\theta\Delta t)^{s+\tau-k} \sigma^{2} \Delta t \\ N(0, 1, (i-1)\Delta t) N(0, 1, (k+s-1)\Delta t \Big] \Bigg\rangle.
    \end{gathered}
    \end{equation}
Using linearity of the expectation value yields
    \begin{equation}
    \begin{gathered}
        \sum_{k=1}^{\tau}\sum_{i=1}^{s}[(1-\theta\Delta t)^{s+\tau-k}\theta\Delta t \sigma \sqrt{\Delta t} \langle N(0, 1, (k+s-1)\rangle]  + \\  \sum_{k=1}^{\tau}\sum_{i=1}^{s}\Big[(1-\theta\Delta t)^{s+\tau-k} \sigma^{2} \Delta t \\ \left\langle N(0, 1, (i-1)\Delta t) N(0, 1, (k+s-1)\Delta t \right\rangle\Big]  .
    \end{gathered}
    \end{equation}
By definition the expectation value of the normal distribution $\langle N(0, 1, t) \rangle = \langle \Gamma(t) \rangle = 0 \rangle$ (Eq. \ref{Eq.Gamma1}) and the auto-correlation $\langle N(0, 1, t)N(0, 1, t+\tau) = \langle \Gamma(t)\Gamma(t+\tau) \rangle = \delta(\tau)$ (Eq. \ref{Eq.Gamma2}). Thus the first summand vanishes and the equation reads
    \begin{equation}
    \sum_{k=1}^{\tau}{\sum_{i=1}^{s}{[(1-\theta\Delta t)^{s+\tau-k} \sigma^{2} \Delta t \delta((k+s-1)\Delta t-(i-1)\Delta t)]  }}
    \end{equation}
$(k+s-1)-(i-1)=k+s-i$ with $k>1$ and $i<s$. Thus $k+s-i>0 \;\forall k, i$. $\rightarrow \delta((k+s-i)\Delta t) = 0 \;\forall k, i$. 
    \begin{equation}
   \Rightarrow\left\langle \sum_{k=1}^{\tau}{x(s) (1-\theta\Delta t)^{\tau-k} \sigma \sqrt{\Delta t} N(0, 1, (k+s-1)\Delta t)} \right\rangle  = 0. 
   \label{Eq.app14dg}
    \end{equation}
\end{subequations}
\rule{\columnwidth}{.5pt}
Inserting Eq. (\ref{Eq.app14dg}) into Eq. (\ref{Eq.app14c}) and applying Eq. (\ref{Eq.mean})
    \begin{equation}
    (1-\theta\Delta t)^{\tau}\left\langle x^{2}(s) \right\rangle + \sum_{k=1}^{\tau}{\left[ (1-\theta\Delta t)^{\tau-k} \theta\Delta t \mu^{2} \right]}.
    \label{Eq.app14e}
    \end{equation}
Considering the relative auto co-variance, between two time steps, following statement must hold
    \begin{equation}
    \langle x(t_{1})x(t_{1}+\tau \rangle \overset{!}{=} \langle x(t_{2})x(t_{2}+\tau \rangle
    \label{Eq.app14f}
    \end{equation}
    \begin{equation}
    \begin{gathered}
    \overset{\text{\ref{Eq.app14e} in \ref{Eq.app14f}}}{\Leftrightarrow} (1 - \theta \Delta t)^{\tau} \langle x^{2}(t_{1}) + \rangle \sum_{k=1}^{\tau}{\left[ (1-\theta\Delta t)^{\tau-k} \theta\Delta t \mu^{2} \right]} = \\
    (1 - \theta \Delta t)^{\tau} \langle x^{2}(t_{2}) + \rangle \sum_{k=1}^{\tau}{\left[ (1-\theta\Delta t)^{\tau-k} \theta\Delta t \mu^{2} \right]}
    \end{gathered}
    \end{equation}
    \begin{equation}
    \Rightarrow \langle x^{2}(t_{1}) \rangle = \langle x^{2}(t_{2}) \rangle.
    \label{eq:vart1t2}
    \end{equation}
If $t_1$ and $t_2$ is sufficiently large, while the difference $t_1-t_2$ is constant, Eq. (\ref{eq:vart1t2}) reads
    \begin{equation}
    \Rightarrow \lim\limits_{t_{1}\rightarrow \infty}{\langle x^{2}(t_{1}) \rangle} = \lim\limits_{t_{2} \rightarrow \infty}{\langle x^{2}(t_{2}) \rangle}.
    \label{Eq.app14i}
    \end{equation}
Using Eq. (\ref{Eq.var-tot-1}) and (\ref{Eq.var-tot-2}), Eq. (\ref{Eq.app14i}) becomes
    \begin{equation}
    \Rightarrow \frac{\sigma^{2}\Delta t}{1-(1-\theta \Delta t)^{2}} = \frac{\sigma^{2}\Delta t}{1-(1-\theta \Delta t)^{2}} \;\square
    \end{equation}. 
\end{subequations}

\begin{acknowledgements}
        We thank Greg Madejski for extensive discussions on blazar variability. We also thank the referee Dr.\,Norris for conscientious comments and constructive criticism. SMW acknowledges support by {\em Stiftung der deutschen Wirtschaft} and hospitality of the KIPAC at SLAC. The {\em DATEV Stiftung Zukunft} is acknowledged for funding  the interdisciplinary data lab {\em DataSphere@JMUW} where most of this collaborative research was borne out.
\end{acknowledgements}

\bibliographystyle{jwaabib}
\bibliography{mnemonic.bib,aa_abbrv.bib,bib.bib}
\end{document}